\documentclass[aps,prx,twocolumn,amsmath,amssymb,superscriptaddress,floatfix,longbibliography,title,toc,titletoc,page,toc,page,header]{revtex4-2}

\usepackage{mathtools}
\usepackage{multirow}
\usepackage{xcolor}
\usepackage{bm}
\usepackage{amsfonts,amssymb,amsmath}
\usepackage{graphicx,dcolumn,bm,xcolor,braket,slashed}
\usepackage{times} 
\usepackage{comment}
\usepackage{array}
\usepackage{textcomp}
\usepackage[normalem]{ulem}
\usepackage{dsfont}

\usepackage[hyperfootnotes=false]{hyperref}

\newcommand{\be}{\begin{eqnarray}}
\newcommand{\ee}{\end{eqnarray}}
\newcommand{\bbm}{\begin{bmatrix}}
\newcommand{\ebm}{\end{bmatrix}}
\newcommand{\bpm}{\begin{pmatrix}}
\newcommand{\epm}{\end{pmatrix}}
\newcommand{\dg}{\dagger}

\makeatletter
\g@addto@macro\normalsize{%
  \setlength\abovedisplayskip{12pt}%
  \setlength\belowdisplayskip{12pt}%
  \setlength\abovedisplayshortskip{4pt}%
  \setlength\belowdisplayshortskip{4pt}%
}
\makeatother

\begin{document}

\title{
Cavity quantum electrodynamics of photonic temporal crystals
}

\author{Junhyeon Bae}
\affiliation{Department of Physics, Chung-Ang University, 06974 Seoul, Republic of Korea}

\author{Kyungmin Lee}
\affiliation{Department of Physics, Korea Advanced Institute of Science and Technology, Daejeon 34141, Republic of Korea}

\author{Bumki Min}
\email{bmin@kaist.ac.kr}
\affiliation{Department of Physics, Korea Advanced Institute of Science and Technology, Daejeon 34141, Republic of Korea}

\author{Kun Woo Kim}
\email{kunx@cau.ac.kr}
\affiliation{Department of Physics, Chung-Ang University, 06974 Seoul, Republic of Korea}

\begin{abstract}
Photonic temporal crystals host a variety of intriguing phenomena, from wave amplification and mixing to exotic band structures, all stemming from the time-periodic modulation of optical properties. While these features have been well described classically, their quantum manifestation has remained elusive. Here, we introduce a quantum electrodynamical model of PTCs that reveals a deeper connection between classical and quantum pictures: the classical momentum gap arises from a localization–delocalization quantum phase transition in a Floquet-photonic synthetic lattice. 
Leveraging an effective Hamiltonian perspective, we pinpoint the critical momenta and highlight how classical exponential field growth manifests itself as wave-packet acceleration in the quantum synthetic space. Remarkably, when a two-level atom is embedded in such a cavity, its Rabi oscillations undergo irreversible decay to a half-and-half mixed state—a previously unobserved phenomenon driven by photonic delocalization within the momentum gap, even with just a single frequency mode. Our findings establish photonic temporal crystals as versatile platforms for studying nonequilibrium quantum photonics and suggest new avenues for controlling light matter interactions through time domain engineering.

\end{abstract}

\maketitle
\newpage
\noindent \textit{Introduction.---}
Recent advances in photonics have moved beyond spatially periodic media such as conventional photonic crystals and metamaterials \cite{galiffi2022photonics,rizza2024harnessing,lee2018linear,solis2021time,caloz2019spacetime,xiao2014reflection,Chamanara2018,Sounas2017}, turning instead to structures whose optical properties are periodically modulated in time \cite{tirole2023double,PhysRevLett.123.206101,Pendry:21,lyubarov2022amplified}. Among these dynamic media are photonic temporal crystals (PTCs), in which parameters such as refractive index or permittivity follow a strict periodic pattern in time \cite{Wang2018photonic,doi:10.1126/sciadv.abo6220,zurita2009reflection,PhysRevA.93.063813,doi:10.1126/sciadv.adg7541,park2024spontaneous,vazquez2023incandescent,asgari2024theory}. This shift to time-domain periodicity has unveiled unusual photonic band structures and unique ways of manipulating light under time-varying conditions \cite{pacheco2020temporal,yin2022floquet,Park2021,Lee2021,wang2024expanding}. PTCs thus offer prospects for advanced control of wave propagation, energy transfer, and frequency conversion, expanding beyond the traditional realm of static spatial lattices.

In classical descriptions, PTCs exhibit non-Hermitian behavior arising from time-periodic modulations, giving rise to gain and loss phenomena whose physical origin is distinct from that of static photonic systems \cite{Pendry:21,park2024spontaneous}. This non-Hermiticity manifests in distinctive band structures marked by momentum gaps and exceptional points, and the breakdown of Floquet-state orthogonality—quantified by the Petermann factor—heavily influences the photonic density of states near the gap edge \cite{Wang2018photonic,doi:10.1126/sciadv.abo6220,miri2019exceptional,wang2019non,park2024spontaneous}. These changes in the photonic density of states are directly relevant for wave amplification, lasing, and frequency conversion, while also influencing atomic light emitter interactions by altering emission and absorption pathways \cite{PhysRevLett.123.206101,park2024spontaneous,doi:10.1126/sciadv.adg7541,doi:10.1126/sciadv.abo6220,Shcherbakov2019}, thus bridging classical PTC phenomena with inherently quantum regimes.

Despite extensive classical analyses via Maxwell’s equations and Floquet theory, a critical gap remains in understanding the quantum properties of PTCs \cite{lyubarov2022amplified}. Classical models can effectively capture non-Hermitian effects (e.g., momentum gaps, exceptional points) but fall short in describing quantum dynamical phenomena unique to time-varying environments. In other words, non-Hermiticity at the classical level can obscure how photons truly behave, particularly when light–matter interactions such as spontaneous emission and Rabi oscillations enter the picture. For instance, the formation of momentum gaps is often attributed to pseudo-Hermiticity breaking, yet quantization of Maxwell’s equations yields a Hermitian Hamiltonian, even in PTCs \cite{lyubarov2022amplified}. This raises a central question: How can the non-Hermitian features observed classically be reconciled with the Hermiticity of quantum electrodynamical descriptions? Clarifying this paradox is vital for exploring purely quantum effects such as photon pair creation, annihilation, squeezing, and Floquet vacuum fluctuations, all of which exceed the scope of classical models.

Building on these foundations, we develop a cavity quantum electrodynamical (QED) model that bridges classical and quantum perspectives of PTCs. Our approach reveals how classical non-Hermitian features, such as momentum gaps and time-periodicity induced gain, map onto a localization-to-delocalization quantum phase transition in a Floquet-photonic synthetic lattice. Notably, we show that Rabi oscillations in a two-level system coupled to a PTC settle into an incoherent mixed state within the momentum gap, even in a single-frequency cavity, a phenomenon triggered by photonic delocalization under time-periodic driving. This mechanism is inaccessible through classical theories alone, placing PTCs at the forefront of non-equilibrium quantum photonics. Our results highlight PTCs as a powerful platform for exploring new light–matter interaction regimes and advancing quantum photonic technologies.
\newline 

\begin{figure*}
    \centering
    \includegraphics[width=0.9\textwidth,clip]{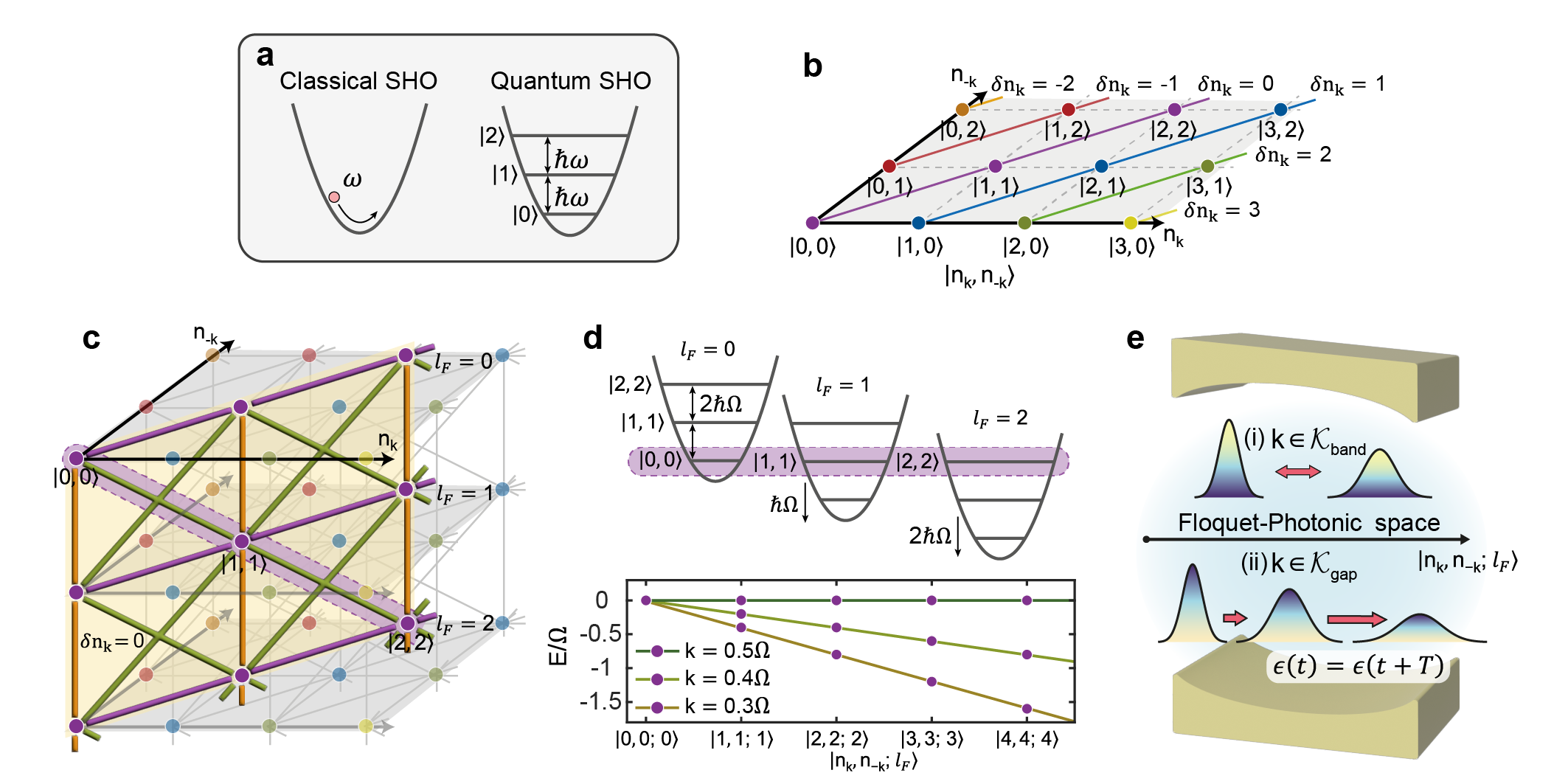}
    \caption{(a) When a classical simple harmonic oscillator (SHO) with frequency $\omega$ is quantized, the energy spacing between neighboring eigenstates is $\hbar\omega$.
    A similar principle applies to the eigenfrequency spectrum of classical PTCs and the eigenenergy spectrum of quantum PTCs. (b) The 2D photon number space of a quantum PTC can be grouped with states that share the same momentum $P = \hbar c k\,(n_k - n_{-k})$, which is a good quantum number. The onsite potential and hopping amplitude are time-dependent, as shown in Eq.~\eqref{eq:Ht2}.  
    (c)  By applying the Floquet formalism to the time-periodic Hamiltonian of a quantum PTC, the photonic lattice originally described by $(n_k, n_{-k})$ acquires a third dimension labeled by the Floquet index $l_F = \dots, -1, 0, 1, 2, \dots$.  (d) At $k = \Omega/2$, the onsite energies of sites connected by pair creation and annihilation coincide, leading to degeneracy. Even a small coupling $\alpha_c$ between these sites can delocalize the eigenstates in the Floquet-photonic space. 
    (e) A schematic depiction of wave-packet dynamics in the Floquet--photonic space: for $k \in \mathcal{K}_{\text{band}}$, the wave packet remains localized and oscillatory, whereas for $k \in \mathcal{K}_{\text{gap}}$, it becomes delocalized and accelerates in the momentum gap.    
    }
    \label{fig1}
\end{figure*}

\begin{figure}
    \centering    \includegraphics[width=0.5\textwidth,clip]{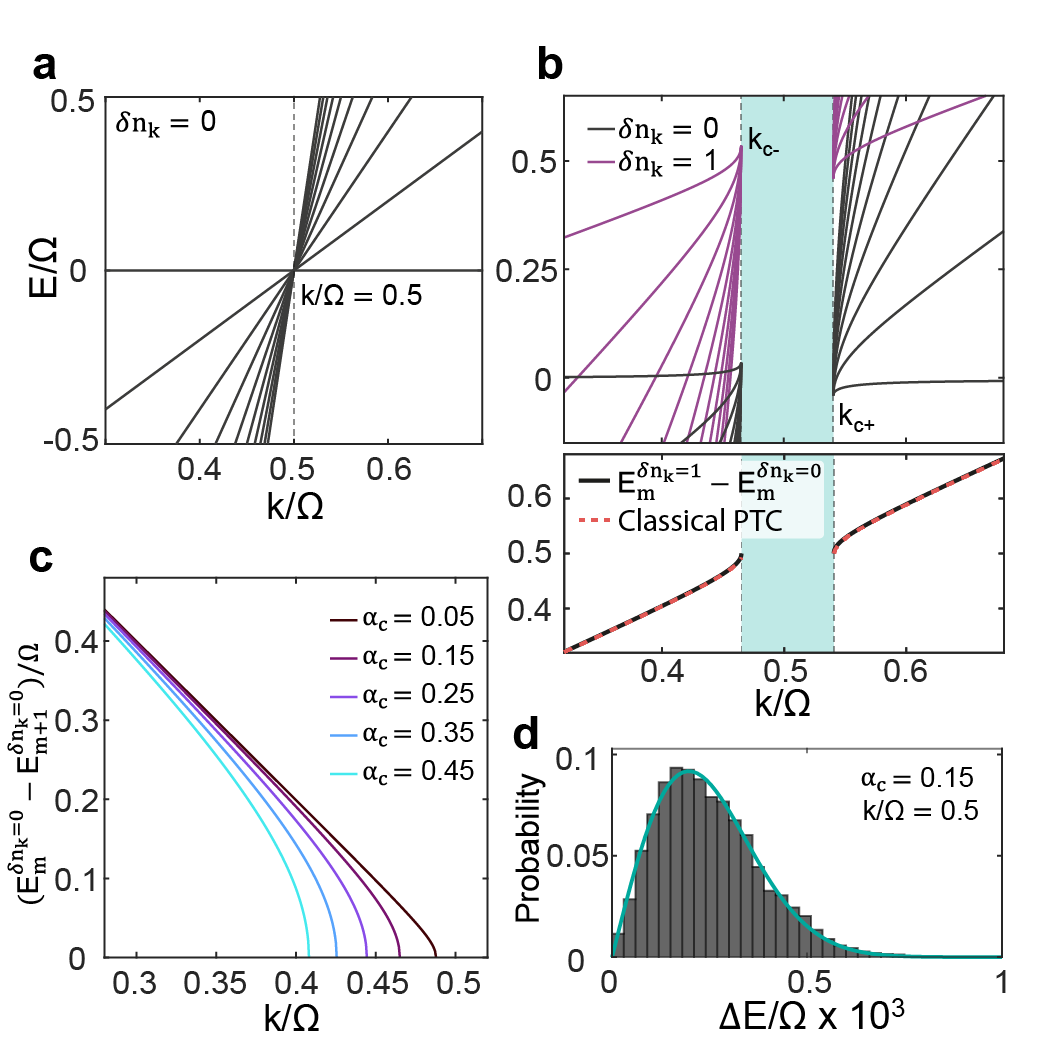}
    \caption{Spectral properties of the quantum PTC. (a) In the absence of driving ($\alpha_c{=}0$), all Floquet-photonic sites $\ket{n_k,n_{\text -k};l_F}$ are decoupled, and their energies are given by $E{=} k (n_k{+}n_{\text{-}k}){-}l_F\Omega$. In this plot, we show the eigenenergy spectrum for $n_k{=}n_{\text{-}k}{=}l_F{=}0,1,2,\cdots$. (b) Eigenenergies of the quantum PTC at $\alpha_c {=} 0.15$ for $\delta n_k {=} 0, 1$. The degeneracy at $k {=} \Omega/2$ is lifted, giving rise to $k_{c\pm} {=} \Omega\,(2 \mp \alpha_c)^{-1}$. Notably, the energy spacing between eigenstates that differ by one momentum quantum, $\bigl(E_m^{\delta n_k=1} {-} E_m^{\delta n_k=0}\bigr)$, precisely reproduces the classical PTC spectrum (see lower panel). (c) For a given $k$, the energy levels of eigenstates that share the same momentum are equally spaced (see Appendix~\ref{appendix:A}), and this spacing converges to zero at the critical momentum $k_{c-}$ (d) For $k \in \mathcal{K}_{\text{gap}}$, the distribution of energy-level spacings follows Wigner--Dyson (orthogonal) statistics, indicating that the eigenstates are delocalized within the momentum gap.}
    \label{fig2}
\end{figure}

\noindent \textit{Effective description of the quantum PTC.---}We consider a cavity that supports a \textit{single}  frequency photonic mode (Fig.~\ref{fig1}), with a time-periodic permittivity $\varepsilon(t){=}\varepsilon(t+T)$. In natural units adapted for electromagnetism ($\hbar{=}c{=}\varepsilon_0{=}1$), the Hamiltonian of the system is given by~\cite{lyubarov2022amplified}
\begin{align}
    \hat H(t)=  k(1-\alpha(t)) (\hat n_k + \hat n_{\text{-}k}) +k\alpha(t) (\hat a_k \hat a_{\text{-}k} + h.c.), 
    \label{eq:Ht}
\end{align}
where $\hat n_k{=}\hat a^\dg_k \hat a_k$ is the photon number operator for mode \(k\), and $\alpha(t){=}\frac{1}{2}\left[1{-}\varepsilon^{-1}(t)\right]$. 
In the photon number basis, defined by $\hat{n}_k \ket{n_k} {=} n_k \ket{n_k}$, the first term of the Hamiltonian corresponds to the energy contribution from the photon number. The second term describes the creation and annihilation of photon pairs, governed by the operator $\hat{a}_k \hat{a}_{\text -k}$, which acts as $\hat{a}_k \hat{a}_{\text -k} \ket{n_k, n_{\text -k}} {=} \sqrt{n_k n_{\text -k}} \ket{n_k {-} 1, n_{\text -k} {-} 1}$. This term generates off-diagonal elements in the Hamiltonian, representing a \textit{hopping} process between adjacent photon number states. As a result, the Hamiltonian can be interpreted as a single-particle problem on a 1-dimensional lattice, with both the onsite potential and hopping amplitudes varying periodically in time. The Hamiltonian is expressed as
\begin{align}
    \hat H(t) = \sum_{\bold n} k\mu_{\bold n}(t) \hat c^\dg_{\bold n} \hat c_{\bold n} + k\alpha_{\bold n}(t) (\hat c^\dg_{{\bold n}}\hat c_{{\bold n}\text{-}\bold 1} +h.c.),
    \label{eq:Ht2}
\end{align}
where $\mu_{\bold n}(t){=}(1{-}\alpha(t))(n_k{+}n_{\text{-}k})$, $c^\dg_{\bold n} \ket{0}{=}\ket{n_k,n_{\text{-}k}}$, and $\alpha_{\bold n}(t){=}\sqrt{n_kn_{\text{-}k}}\alpha(t)$. Hereafter, we adopt the specific form $\alpha(t){=}\alpha_c\cos (\Omega t)$ for the time-periodic modulation of the permittivity. The coefficients depend not only on time, but also on the \textit{photonic site} index, $\bold n {=}(n_k,n_{\text{-}k})$. The lattice sites are defined only for non-negative photon numbers, $n_{k},n_{\text -k}{\ge} 0$. The linearly increasing onsite potential, known as the Wannier-Stark ladder~\cite{wanni1960wave,emin1987exis,avro1994perio,wilk1996obser,gluck2002wannier}, leads to the localization of eigenstates. On the other hand, the hopping amplitude varies linearly across the photonic sites, establishing the quantum PTC as a promising new model system for investigating unexplored phenomena. Later, we show that this enables the system to undergo a localization-to-delocalization transition, which occurs at the edges of the momentum gap. Transformed to a time-periodic single-particle problem, the Floquet formalism is employed to describe it in the frequency domain:
\begin{align}
    (\hat H_F)_{l_Fl_F'}=-l_F\Omega \hat \delta_{l_Fl_F'}+\frac 1 T \int_0^T \hat H(t) e^{-i\Omega(l_F-l_F')t}dt.
    \label{eq:floquettransform}
\end{align}
The Hilbert space is spanned by the photon number basis and the Floquet basis, both of which are infinite-dimensional, see Fig.~\ref{fig1}c. While the Floquet Hamiltonian is in the 3-dimensional synthetic space \((n_k, n_{\text{-}k}, l_F)\), it can be simplified to a 1-dimensional form by collecting Floquet-photonic sites that are relevant only for the formation of the quantum momentum gap near \(k {=} \Omega/2\) and \(E {=} 0\). For instance, the synthetic sites are grouped by the blue dotted line in Fig.~\ref{fig1}c, and they are coupled by the time-periodic modulation of the permittivity.
Approaching $k{=}\Omega/2$, the potential energies of the sites become increasingly similar, as shown in Fig.~\ref{fig1}d. Consequently, the effective physics can be described by the following effective Floquet Hamiltonian when $\alpha_c \ll 1$:
\begin{align}
    \hat H_F^{\text{eff}} = \sum_{n=0}^\infty \left(k-\frac \Omega 2 \right)n\hat c^\dg_n \hat c_n+\alpha_c n (\hat c^\dg_n\hat c_{n-1}+h.c.),
    \label{eq:HF}
\end{align}
where the Floquet-photonic basis $n$ is introduced, with $n_k{=}n_{\text -k}{=}l_F{=}n$. The momentum $P{=} (n_k{-}n_{\text - k}) k$ is a good quantum number, and for simplicity, we focus on the zero-momentum sector with $n_k{=}n_{\text{-}k}$. The underlying physics for $\delta n_k {=}n_k {-} n_{\text{-}k}{\neq} 0$ remains consistent and unaffected (see Appendix~\ref{appendix:A}). When \(\alpha_c {=} 0\) (i.e., not driven), all (quasi-)eigenenergies of $\hat H_F^{\text{eff}}$ with $\delta n_k{=}0$ are degenerate at zero energy at \(k {=} \Omega/2\) (see Fig.~\ref{fig2}a). Remarkably, 
this degeneracy persists even with a nonzero \(\alpha_c\), which shifts the critical momentum (i.e., the edge of momentum gap) to \(k_{c\pm} {=} \Omega/2 + \delta k_{c\pm}\) (see Fig.~\ref{fig2}b). Note that eigenenergy $E^{\delta n_k}_m$ is indexed by $m{=}1,2,\cdots$ in ascending order of the photon number expectation value $\langle \hat N \rangle$ for each eigenstate $\ket{\psi_m}$ sharing the same momentum $\delta n_k$.
\newline

\noindent \textit{The quantum vs classical momentum gap edge.---} 
The aforementioned effective Floquet Hamiltonian has no upper or lower bound, with its elements growing indefinitely as the photon number and Floquet index increase. When diagonalized, the eigenstates converge to a single energy as $k \rightarrow k_{c\pm}+0^{\pm}$, thereby making it challenging to accurately estimate $k_{c\pm}$ using brute-force numerical diagonalization with truncation. We circumvent the issue by employing the transfer matrix method \cite{Ishii1973Localization, mackinnon1983scaling, J.L.Pichard_1981Finitesizescaling, crit2009slev,Dwivedi2016generalized, Luo2021Transfermatrix,Zhang2022Lyapunovexponent, anis2023xiao} for the effective Floquet Hamiltonian. 
The transfer matrix provides a recursive relationship between consecutive quantum states in the Floquet-photonic space: 
$\Psi{(n)} {=} T_n \Psi{(n-1)}$, where $\Psi(n) {=}[\psi{(n+1)},\, \psi{(n)}]^T $ 
(see Appendix~\ref{appendix:B}). At the critical momentum, the norms of the transfer eigenvalues become unity as the energy gap between the modes vanishes. Using this condition, one can identify the values of the critical momenta, which, in the large $n$ limit, become independent of energy and momentum. This robustly defines the quantum momentum gap.
As a result, the critical momenta are given by
\begin{align}
    k_{c-} = \frac{\Omega}{2+\alpha_c},\,\,\,\,\,\,\,\,k_{c+} = \frac{\Omega}{2-\alpha_c},
\end{align}
when the permittivity is expressed as $\varepsilon^{-1}=1-2\alpha_c \cos(\Omega t)$. This represents the exact result, and the method can be extended to a more general time-periodic permittivity (see Appendix~\ref{appendix:C}). The location of the momentum gap edge in the classical PTC can also be determined to arbitrary precision using the transfer matrix approach (see Appendix~\ref{appendix:C}). 
Since the momentum gap $\mathcal K_{\text{gap}} {=}\{ k| k_{c-}{<}k{<}k_{c+}\}$ is formed by the hybridization of an infinite number of photonic modes sharing the same momentum, the classical description
precisely predicts the location of \(k_{c\pm}\) without any quantum corrections. This is numerically verified by comparing the eigenvalues of \(\hat{U}_F {=} \mathcal{T} \exp \left(-i \int_0^T \hat{H}(t) dt \right)\) with those of the classical PTC. When \(\hat{H}^{\text{eff}}_F\) is used, it introduces only a constant shift in the critical momenta, preserving all relative features near the critical momentum (see Appendix~\ref{appendix:A} and~\ref{appendix:D}). This implies that the physics near \(k_{c\pm}\) is accurately captured by the effective Floquet Hamiltonian in Eq.~\eqref{eq:HF}.

The momentum gap transition in the classical PTC is reflected in the imaginary part of the \textit{eigenfrequencies}, 
leading to the exponential growth of the fields. In contrast, the quantum Hamiltonian is Hermitian, ensuring that all its eigenvalues are real. The transition in the quantum PTC manifests as a localization-to-delocalization transition of eigenstates in the Floquet-photonic lattice, which is reflected in the distribution of eigenenergy level spacings. For $k{<}k_{c-}$, the eigenenergies are equally spaced (Fig.~\ref{fig2}a,b). In contrast, for $k\in \mathcal K_{\text{gap}}$, the statistics of eigenenergy level spacings follow the Wigner-Dyson distribution (orthogonal class)~\cite{c5bd8f0f-2576-3f83-a184-791e55682183,mehta2004random} (see Fig.~\ref{fig2}d), indicating that the eigenstates are delocalized in the Floquet-photonic space. We next examine the transport properties using both the transfer matrix approach and the numerical time evolution of a wave packet in both regimes. This approach offers a unified perspective on both the photonic band structure and the momentum gap, framed within the context of the localization-to-delocalization transition. Furthermore, it interprets the physics of the classical PTC in terms of its underlying microscopic Hamiltonian. 
\newline 

\noindent \textit{The localization-to-delocalization transition.---} 
The effective Floquet Hamiltonian in Eq.~\eqref{eq:HF} captures two competing mechanisms that govern the dynamical behavior of a wave packet. The first term represents an onsite potential that increases (or decreases) linearly with the site index for \(k {<} \Omega/2\) (or \( k {>} \Omega/2\)). This term induces a Wannier-Stark effect, which localizes eigenstates in systems with finite bandwidth~\cite{emin1987exis,avro1994perio,kim2016surface,anom2020kim}. The second term describes a typical hopping Hamiltonian, but with a hopping amplitude that also increases with the site index, thereby promoting delocalization of quantum states in the Floquet-photonic space. Together, the increasing kinetic and onsite energies create a new quantum mechanical setting, where a localization-to-delocalization transition emerges at the critical momentum \(k_{c\pm}\) in a PTC, even in the absence of quenched disorder.

\begin{figure}
    \centering    \includegraphics[width=0.48\textwidth,clip]{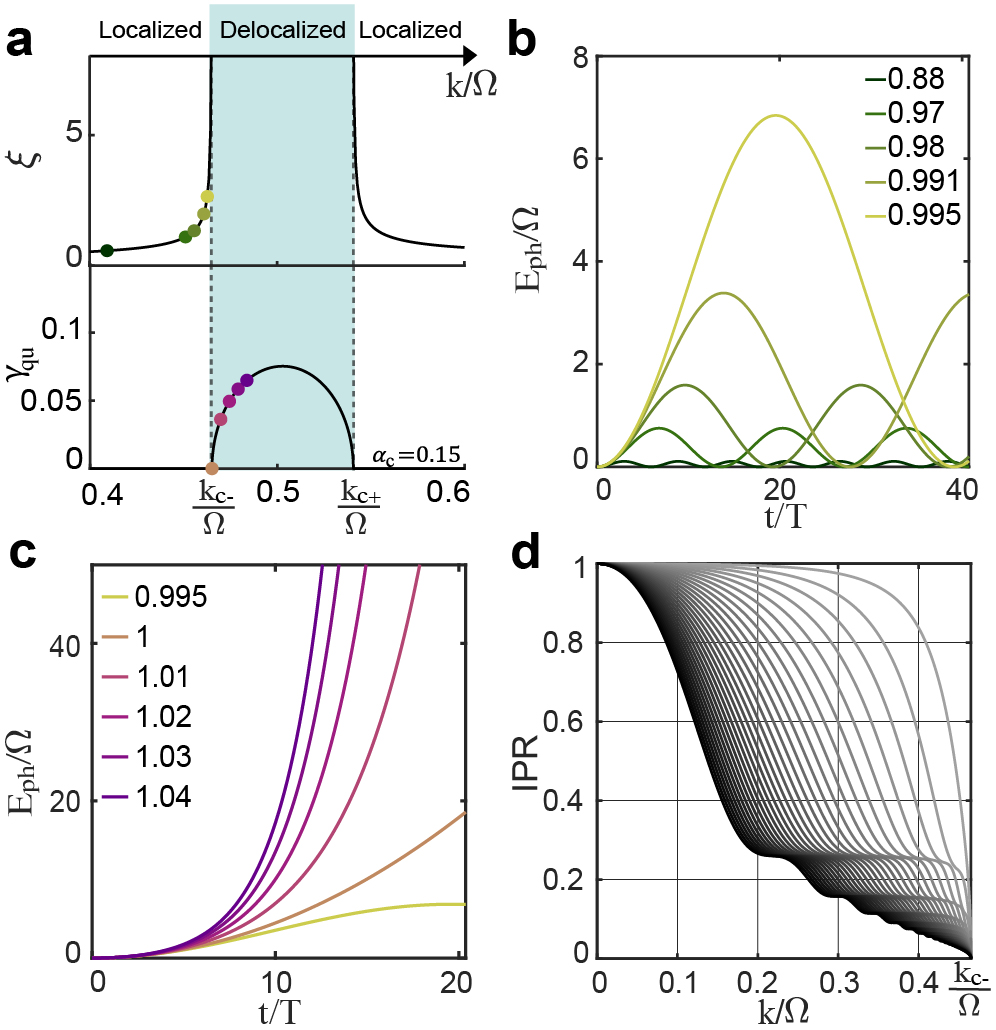}
    \caption{Localization-to-delocalization transition in the Floquet-photonic synthetic space. (a) The localization length $\xi$ [Eq.~\eqref{eq:loc}] in the band (upper panel) and the photonic-energy growth rate [Eq.~\eqref{eq:gamma}] in the momentum gap (lower panel). (b) Oscillatory photonic energy $E_{\text{ph}} = k \langle \hat{N} \rangle$ over time for $k < k_{c-}$, where the initial wave packet is prepared at $n_k{=}n_{\text{-}k}{=}l_F{=}0$. Colors in the legend indicate the ratio $k/k_{c-}$. (c) For  $k\in \mathcal K_{\text{gap}}$, the photonic energy of the same initial wave packet grows exponentially over time. (d) The inverse participation ratio of eigenstate $\ket{\psi_m}$ ($m{=}0,\cdots,50$) with $\delta n_k{=}0$.}
    \label{fig3}
\end{figure}

An essential measure that determines the (de)localization of quantum states is the Lyapunov exponent, defined as  
\begin{align}
\gamma = \lim_{N \rightarrow \infty} \frac{1}{N} \ln\left[\frac{|\Psi(n{=}N)|}{|\Psi(n{=}0)|}\right], 
\end{align}  
where $\Psi(n)= T_n T_{n-1}\cdots  T_1 \Psi(0)$. $\gamma$ converges to zero for a delocalized state (indicating that the norm of the state at site 0 is equal to that of the state infinitely far away) or to a positive finite value for a localized state, $ |\psi(n)| \sim e^{-\gamma n} $ (see Appendix~\ref{appendix:B}). The localization length, \(\xi = 1/\gamma\), is calculated both analytically and numerically for the effective Floquet Hamiltonian, Eq.~\eqref{eq:HF}, of the quantum PTC (see Fig.~\ref{fig3}a). It fully characterizes the nature of the transition at the quantum momentum edge (see Appendix~\ref{appendix:B}).   For $k < k_{c-}$, the localization length is given by  
\begin{align}
     \xi^{-1} = \text{arcosh} \left( \frac{1-\frac \Omega{2k}}{1-\frac{\Omega}{2k_{c-}}}\right),
     \label{eq:loc}
\end{align}
which diverges near the momentum gap edge as $\xi \sim |k - k_{c-}|^{-\nu}$ with the critical exponent $\nu = 1$. For $k\in \mathcal K_{\text{gap}}$, the localization length $\xi$ diverges, $\xi\rightarrow\infty$. Figure~\ref{fig3}b provides a numerical demonstration of this result. The height of the oscillations indicates the extent to which a wave packet can spread in the Floquet-photon space and is proportional to the localization length. The periodicity of the oscillations can be deduced by the energy differences between neighboring eigenstates comprising the initial wave packet at $\ket{0,0;l_F{=}0}$ (see Appendix~\ref{appendix:F}), which diverge as the critical momentum is approached.

\begin{figure*}
    \centering    \includegraphics[width=0.95\textwidth,clip]{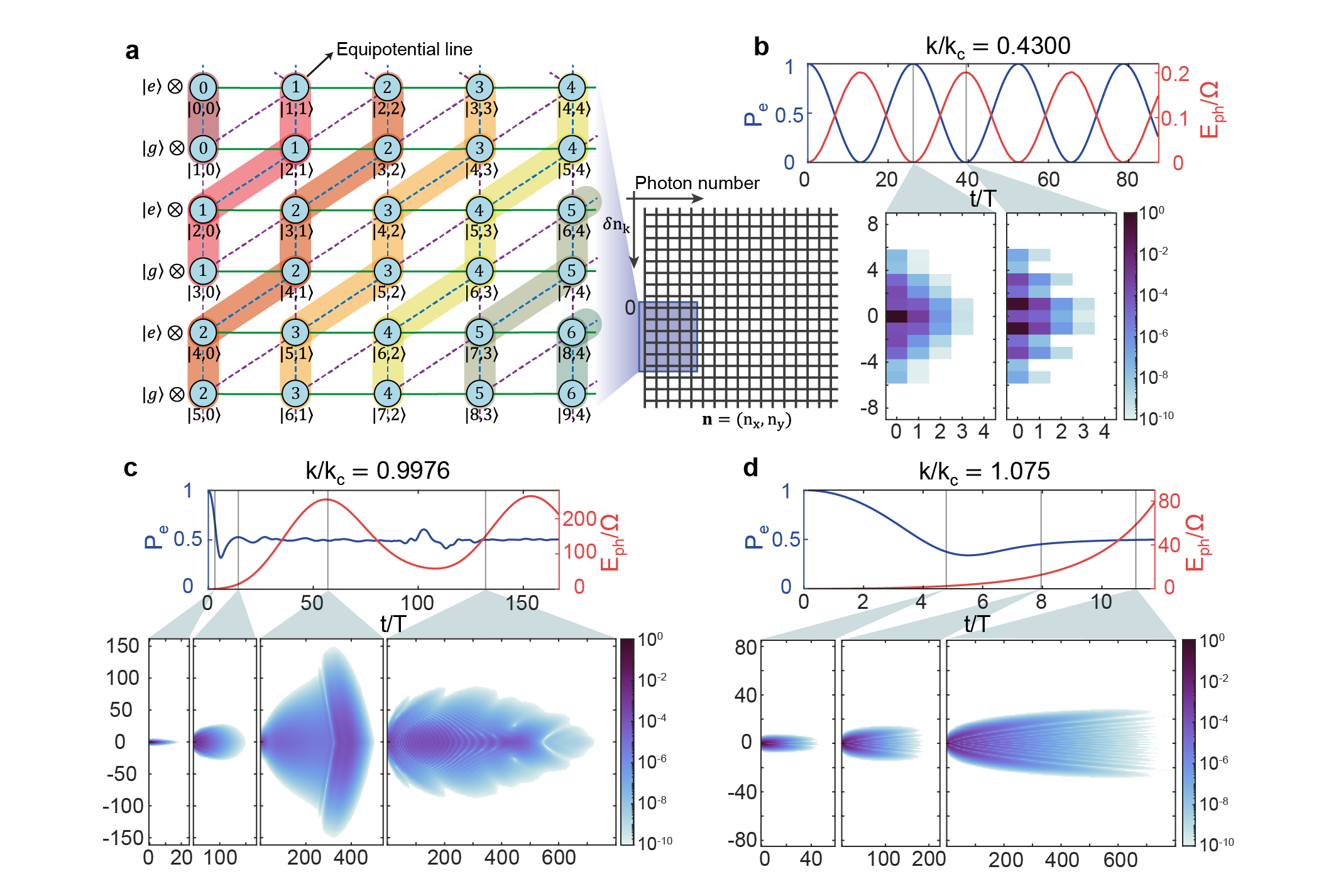}
    \caption{Dynamics of the quantum Rabi model in the PTC. (a) Schematic of the expanded Floquet-photonic lattice $\boldsymbol{n}$ combined with atomic states, indicated on the left end of each line. The Floquet index $l_F$ and the photon number state $\ket{n_k,n_{\text - k}}$ are marked within and next to each site, respectively. When the atomic energy is set to $\Delta {=} k$, all onsite energies among the illustrated lattice sites coincide at $k {=} \Omega/2$. At $k{\neq} \Omega/2$, sites with identical onsite energies are grouped together by color. Neighboring sites are coupled by $\alpha_c$ (green dotted lines), $g$ (blue dotted lines), and $\alpha_c g$ (purple dotted lines). (b) At $k \ll k_{c-}$, both the excited-atomic-state population and the photonic energy oscillate coherently, akin to the conventional quantum Rabi model. In the lower panel, $|\psi_t(\mathbf{n})|^2$ on the lattice shown in (a) is plotted (log scale) at $t/T = 160$ and $t/T = 250$. The initial wave packet at $t=0$ is prepared in $\ket{0,0;0} \otimes \ket{e}$, positioned at $\mathbf{n} = (0,0)$. (c) At $k/k_{c-} {=} 0.9976$, the initial atomic state $\rho_{\text{at}}(t{=}0) {=} |e\rangle \langle e|$ decays into the mixed state $\rho_{\text{at}} = \tfrac12 \bigl( |g\rangle \langle g| + |e\rangle \langle e| \bigr)$ because the wave packet spreads over many Floquet-photonic sites with a large localization length ($\xi \gg 1$). Nonetheless, the photonic energy retains its oscillatory behavior. (d) For $k \in \mathcal{K}_{\text{gap}}$, the photonic energy diverges exponentially as the atomic state decays. The lower panel illustrates continuous propagation of the quantum states, showing no sign of localization.}
    \label{fig4}
\end{figure*}

Within the momentum gap ($k {\in} \mathcal{K}_{\text{gap}}$), an initially prepared wave packet at the Floquet-photonic site $n {=} 0$, $\ket{\psi_{t{=}0}} {=} \ket{0,0; l_F {=} 0}$, not only spreads to higher nonzero photonic number states but also accelerates over time. 
The acceleration of the wave packet in the Floquet–photonic space is associated with the rate of energy transfer from the PTC, where the total number of photons increases as energy is drawn from the source driving the PTC.
The quantification of the acceleration is numerically shown in Fig.~\ref{fig3}c and can be analytically understood by calculating the time derivative of $\langle \hat N \rangle {=} \bra {\psi_t} \hat N \ket {\psi_t}$, i.e., the center of probability in the Floquet-photonic lattice, where $\ket {\psi_t}{=}e^{-iH^{\text{eff}}_F t}\ket{\psi_{t{=}0}}$:
\begin{align}
     \frac{d \langle \hat N \rangle }{dt} 
     &{=}\frac{1}{i}\bra {\psi_t} [\hat H_F^{\text{eff}},\hat N] \ket{\psi_t} {=} \sum_{n=0}^\infty n \alpha_c k \Im [\psi_t^*(n) \psi_{t}(n+1)],
\end{align}
which quantifies the photonic energy current flowing into the system. The relation $\psi_t(n{+}1) {=} e^{i\phi} \psi_t(n)$ can be applied as the eigenstates are delocalized ($\xi {\rightarrow} \infty$), where $\phi {=} \arccos \left( \frac{1 - \Omega/2k}{1 - \Omega/2k_{c-}} \right)$ in the large-$n$ limit (see Appendix~\ref{appendix:B}). On the other hand, the average number of photons in the system is given by $\langle \hat{N} \rangle {=} \sum_n n[\psi_t^*(n) \psi_t(n)]$. Consequently, the exponential growth rate of photonic energy 
is  
\begin{align}
    \gamma_{\text{qu}} = \frac{1}{\langle \hat{N} \rangle} \frac{d \langle \hat{N} \rangle}{dt} = 
    2k \sqrt{\left( \frac{\Omega}{2k_{c-}} - 1\right)^2 - \left(\frac{\Omega}{2k} - 1\right)^2}, 
    \label{eq:gamma}
\end{align}  
which is exactly twice the imaginary part of the (quasi-)eigenfrequency within the momentum gap in the classical PTC, $\gamma_{\text{qu}} = 2\gamma_{\text{cl}}$.

In the classical PTC, the eigenfrequencies acquire imaginary parts within the momentum gap, resulting in an exponential growth of the field strength with an exponent $\gamma_{\text{cl}}$. Consequently, the energy growth rate of electromagnetic waves is $2\gamma_{\text{cl}}$, which is in agreement with the quantum calculation. From a physical perspective, the local kinetic bandwidth ($\sim \alpha_c n$) increases with $n$, leading to a corresponding increase in the effective group velocity. Because the eigenstates are delocalized for $k_{c-} {<} k {<} k_{c+}$, the wave packet extends into the large-$n$ region, where it acquires a higher group velocity for propagation. The energy required for the acceleration is supplied by the external driving force of the quantum PTC, $\varepsilon(t)$.

Lastly, the inverse participation ratio of the eigenstate $\ket{\psi_m}$, $\text{IPR}_{\psi_m}{=}\sum^N_{n=0}|\psi_m(n)|^4$, another essential measure of the localization-to-delocalization transition, in the Floquet-photonic space is plotted in Fig.~\ref{fig3}d. As the system approaches the critical momentum ($k{\rightarrow} k_{c-}$), the IPR converges to zero in the thermodynamic limit ($N{\rightarrow}\infty$), which is consistent with the diverging localization length. The plateaus of the IPR, one of the unique features in the quantum PTC, arise in the process of delocalization of eigenstates toward $n{=}0$ direction (see Appendix~\ref{appendix:E} for details). Next, we present Rabi oscillations in the quantum PTC, a distinguishing feature of purely quantum behavior.
\newline

\noindent \textit{The dynamics of quantum Rabi model in the  PTC.---}  
When a two-level atom with an electric dipole moment is placed inside a PTC \emph{cavity}, its interaction with the quantized electromagnetic field is governed by the well-known quantum Rabi model ~\cite{PhysRev.49.324,braak2011integrability,xie2017quantum}. This model predicts coherent oscillations of the ground- and excited-state populations, driven by their coupling to the cavity mode. The dynamics of the two-level system is described by the Hamiltonian $\hat{H}_{\text{at}} {=} \Delta (\hat{I} + \hat{\sigma}_z)/2$, expressed in the eigenbasis of the ground and excited states. The corresponding electric dipole moment operator is $\hat{d} {=} d_0 \hat{\sigma}_x$. 
In a single-frequency PTC cavity, photons with wavenumbers $k$ and $-k$ are coupled to the electric dipole moment of a two-level atom. The interaction Hamiltonian is given by~\cite{milonni2019introduction}:
\begin{align}
    \hat H_{\text{int}} =  i g \sqrt{k}\varepsilon^{-1} \hat \sigma_x \otimes (\hat a^\dg_{k}-\hat a_{\text{-}k})+h.c., 
\end{align}
where the coupling depends on the time-periodic permittivity $\varepsilon(t)$ and is proportional to $\sqrt{k} \sim 1/\sqrt{\lambda}$. The coupling constant $g$ includes additional factors that are independent of both time and momentum. The interaction term facilitates coupling between Floquet-photonic states differing by one momentum unit and the atomic states. Previously, the Hilbert space of the quantum PTC was effectively described by a 1-dimensional wire in the Floquet-photonic space at a fixed momentum. When coupled to atomic states, the Hilbert space accessible to a wave packet expands to an \textit{array} of coupled wires corresponding to different momenta $P {=} k(n_k {-} n_{\text -k})$ (see Fig.~\ref{fig4}a).
When the energy of the excited atomic state is tuned to $\Delta {=} k$, the Floquet-photonic sites combined with the atomic state (shown on the left of Fig.~\ref{fig4}a) all share the same onsite energy at $k {=} \Omega/2$, indicating that the effective physics near $E{=}\Omega/2$ is well captured by those states. Neighboring sites are then coupled by $\alpha_c$, $g$, and $\alpha_c g$. Consequently, all eigenstates can become fully delocalized in the two-dimensional Hilbert space, giving rise to a possible quantum dynamical phase transition in the atom–quantum PTC hybrid system (see Appendix~\ref{appendix:G} for a different choice of atomic energy).

For $k {\ll} k_{c-}$, the eigenstates are well localized, and their coupling to atomic states results in conventional Rabi oscillations (Fig.~\ref{fig4}b). Near $k_{c-}$ and within the momentum gap, their behavior deviates significantly. 
The spreading of an initially localized wave packet leads to irreversible changes in the atomic states, resulting in dissipation phenomena, as shown in Fig.~\ref{fig4}c,d, despite the system being limited to a single-frequency cavity and a single atom. 
The initially excited atomic state exhibits underdamped oscillations before settling into the steady-state mixed state $\rho_{\text{at}} {=} \frac{1}{2} (|g\rangle\langle g| + |e\rangle\langle e|)$, reflecting dissipation caused by coupling to infinitely many Floquet–photonic states in the momentum gap. Interestingly, the steady state that emerges is not the atomic ground state but a half-and-half mixed state (HHMS). This occurs because the eigenstates of the atom–quantum PTC hybrid system are delocalized along both the vertical and horizontal axes in Fig.~\ref{fig4}a. Consequently, the atomic ground and excited states end up equally populated.
The decay rate from the initial state (either excited or ground) to the HHMS can be approximately determined using the Fermi-Golden Rule, provided that the coupling strength between the atom and the quantum PTC is much larger than the energy-level spacing between photonic states~\cite{PhysRevLett.129.140402}. This condition is satisfied within the momentum gap when the infinite-dimensional photonic and Floquet spaces are considered.
Furthermore, it is worth noting that the atomic states are entangled with the momentum parity of the Floquet–photonic states: the excited atomic state is entangled with photonic states of even momentum ($|\delta n_k| {=} 0, 2, 4, \cdots$), while the ground atomic state is entangled with photonic states of odd momentum ($|\delta n_k| {=} 1, 3, 5, \cdots$). Therefore, measuring the atomic state simultaneously projects the Floquet–photonic states.
These phenomena are unique to quantum PTCs and have no direct analogs in classical PTCs. Specifically, we propose that atomic-state dissipation, a distinct feature of quantum PTCs, can be systematically observed within the momentum gap as a function of coupling strength and permittivity modulation. 
\newline

\noindent \textit{Discussion.---}
Spontaneous emission of a photon, accompanied by the de-excitation of an atomic state, occurs when a continuum photonic spectrum is available. That is, the initial quantum state $\ket{n_k {=} 0} \otimes \ket{e}$ transitions to $\ket{n_k {=} 1} \otimes \ket{g}$ when $|\hbar c k - \Delta| \lesssim g$. The entropy associated with a single photon in free space is significantly greater than that of no photon. Consequently, spontaneous emission decay is an irreversible process that dissipates the energy of the excited atomic state. This energy is irretrievably transferred to the vast continuum of photonic modes.
In the quantum PTC, a similar dissipative process occurs but is confined to a single-frequency photonic cavity with a time-periodic permittivity. Near the critical momentum and within the momentum gap, the large, effectively unbounded set of Floquet–photonic states acts as a continuum, allowing the atom to relax irreversibly into a mixed state. It is also worthwhile to note that, because of the delocalization of eigenstates in the atomic-photonic two-dimensional domain (Fig.~\ref{fig4}a), a transition to the mixed state can occur irrespective of the specific form of the initially prepared quantum state. With energy transferred from the driven permittivity to photonic energy, another remarkable phenomenon arises in a nearly symmetric manner: a quantum state $\ket{n_k {=} 0, n_{\text -k} {=} 0} \otimes \ket{g}$, initially prepared in the atomic ground state, undergoes irreversible excitation to the same mixed state (see Appendix~\ref{appendix:H2}).  

In this work, we provide a comprehensive understanding of the classical PTC near the critical momentum and within the momentum gap, leveraging the framework of cavity quantum electrodynamics. While our analysis has focused on weak coupling and a closed quantum system with a single atomic light emitter, future work could explore regimes where stronger coupling or multiple atoms introduce collective phenomena analogous to superradiance. Additional complexity may arise if the cavity is allowed to exchange energy with an external reservoir—potentially stabilizing different steady states or enabling new non-equilibrium phase transitions. We anticipate that circuit QED platforms and rapidly tunable photonic cavities could offer practical testbeds, where time-domain engineering of the optical properties would systematically probe the momentum gap and measure the emerging half-and-half mixed atomic state.
\newline

\noindent \textit{Acknowledgement.---} J.H.B. and K.W.K. are supported by the National Research Foundation of Korea (NRF) grant funded by the Korean government(MSIT) (No.  2020R1A5A1016518). K.L. and B.M. are supported by the National Research Foundation of Korea (NRF) through the government of Korea (NRF-2022R1A2C301335313) and the Samsung Science and Technology Foundation (SSTF-BA2402-02).

\bibliography{reference.bib}


\onecolumngrid

\newpage  
\appendix


\counterwithin{figure}{section}

\section{\,\,\,\,\,\, Quasi-eigenenergy spectra of $\hat U_F$, $\hat H_F$, and $\hat H^\text{eff}_F$ }
\label{appendix:A}

\noindent The Hamiltonian of the quantum PTC is given by~\cite{lyubarov2022amplified}:
\begin{equation}
\hat H(t) = \frac{k}{2} \left(1 + \frac{1}{\varepsilon(t)}\right) \left(\hat a_k^\dagger \hat a_k + \hat a_{-k}^\dagger \hat a_{-k}\right) + \frac{k}{2} \left(1 - \frac{1}{\varepsilon(t)}\right) \left(\hat a_k^\dagger \hat a_{-k}^\dagger + \hat a_k \hat a_{-k}\right),
\end{equation}
where \(\varepsilon(t)\) is a periodic function with period \(T\), i.e., \(\varepsilon(t) = \varepsilon(t + T)\).
The momentum operator is $\hat P= k (\hat n_k - \hat n_{-k}) =  k \delta \hat n_k$ and it commmutes with the Hamiltonian, $[\hat H(t), \hat P]=0$. 
We can determine the energy spectrum  of the Hamiltonian by introducing the Floquet operator $\hat U_F\equiv 
\hat U(T,0)=\mathcal T \exp \left(\ -i\int_0^T \hat H(t)dt\right)$, which governs the time evolution of a state for one period of time, 
\begin{equation}
\ket{\psi(T)} = \hat U_F |\psi(0)\rangle. 
\end{equation}
The Floquet operator can be numerically computed  in the photon number basis by the discretization of time, then eigenstates and eigenvalues can be obtained, $\hat U_F \ket{\psi_m} = \exp\left( -i E^{\delta n_k}_m \, T \right) \ket{\psi_m}$, where $m$ is eigenindex ($m=0,1,2,\cdots$) in the increasing order of photon number expectation value, $\langle \hat N \rangle_{\psi_m} = \bra{\psi_m}\hat N \ket{\psi_m}$.

In the main text we employ the Floquet Hamiltonian formalism, a framework particularly well-suited for analyzing periodically driven systems. This approach allows for a systematic calculation of quasi-eigenenergies by expressing the time-dependent Hamiltonian in terms of its Fourier components. The Floquet Hamiltonian $\hat{H}_F$ is defined through its matrix elements as:
\begin{equation}
(\hat{H}_F)_{l_Fl_F'} = -l_F  \Omega \, \hat \delta_{l_Fl_F'} + \frac{1}{T} \int_0^T \hat{H}(t) \, e^{-i \Omega (l_F - l_F') t} \, dt,
\end{equation}
where $\Omega = 2\pi / T$ is the driving frequency, and $l_F, l_F'$ are the Floquet indices corresponding to the Fourier components of the time-periodic Hamiltonian $\hat{H}(t)$. To simplify the analysis near the critical momentum $k_{c-}$, where the eigenvalues coalesce into a single point (as shown in Fig.~A.1), $\hat{H}_F$ is reduced to an effective Hamiltonian $\hat{H}_F^{\text{eff}}$ by collecting states with the same onsite potential at $k=\Omega/2$ in the Floquet-photonic space. The reduction is performed in the basis $\{|n\rangle_{\delta n_k} \otimes |l_F {=} n\rangle\}$, where $\ket {n}_{\delta n_k} {=} \ket{n_k{=}n+\delta n_k,n_{-k}{=}n}$,  $n = 0, 1, 2, \dots$ denotes the photon occupation number with wave number $-k$, and $l_F$ is the Floquet index. Note that without the loss of generality $l_F=n$ is chosen because Floquet spectrum of $l_F=n+1$ is identical up to a energy shift by $-\Omega$.  The matrix elements of the effective (reduced) Floquet Hamiltonian carrying momentum $P=\delta n_k  k$ are expressed as:
\begin{equation}
(\hat H_F^{\text{eff}})_{n,n';\delta n_k} = -n \Omega\, \delta_{n n'} + \frac{1}{T} \int_0^T \langle n  | \hat H(t) | n' \rangle_{\delta n_k}\, e^{-i \Omega (n - n') t}\, dt.
\end{equation}
For $\varepsilon^{-1}(t) = 1 - 2\alpha_c \cos(\Omega t)$, the reduced Floquet Hamiltonian takes the form (see Eq.\eqref{eq:HF}):
\begin{equation}
\hat H_F^{\text{eff}} = \hat H_{\text{ph}} + \hat H_{\text{hopping}} + \hat H_{\text{Floquet}},
\end{equation}
where the elements are defined as:
\begin{itemize}
    \item {The first diagonal term is from the counting of the number of photons:}
    \begin{equation}
    (\hat H_{\text{ph}})_{n,n;\delta n_k} =  k (2n + \delta n_k),
    \end{equation}
    \item {The second term is the Floquet hopping term between the nearest neighbor in the Floquet-photonic lattice:}
    \begin{equation}
    (\hat H_{\text{hopping}})_{n,n+1;\delta n_k} = (\hat H_{\text{hopping}})_{n+1,n;\delta n_k} = \frac{ k \alpha_c}{2} \sqrt{(n + \delta n_k + 1)(n + 1)},
    \end{equation}
    \item {The third term is diagonal Floquet onsite energy:}
    \begin{equation}
    (\hat H_{\text{Floquet}})_{n,n;\delta n_k} = - n  \Omega.
    \end{equation}
\end{itemize}
All other elements of the matrix are zero. By diagonalizing $\hat H_F^{\text{eff}}$, we obtain the quasi-eigenenergies, which are shown as solid lines in Fig.~A.1. The quasi-eigenenergies obtained from the diagonalization of $\hat U_F$ are plotted in the dashed lines. The two spectra are identical up to a momentum shift as shown in Fig.~A.1(b),
demonstrating the robustness and reliability of the effective Hamiltonian approximation. This indicates that the effective Floquet Hamiltonian captures the essential physics near the critical point. The overall momentum shift between the two spectra can be analytically estimated within an arbitrary precision of $\alpha_c$, see Appendix~\ref{appendix:D}.

\begin{figure}[htb!]
  \centering
    \includegraphics[width=0.9\textwidth]{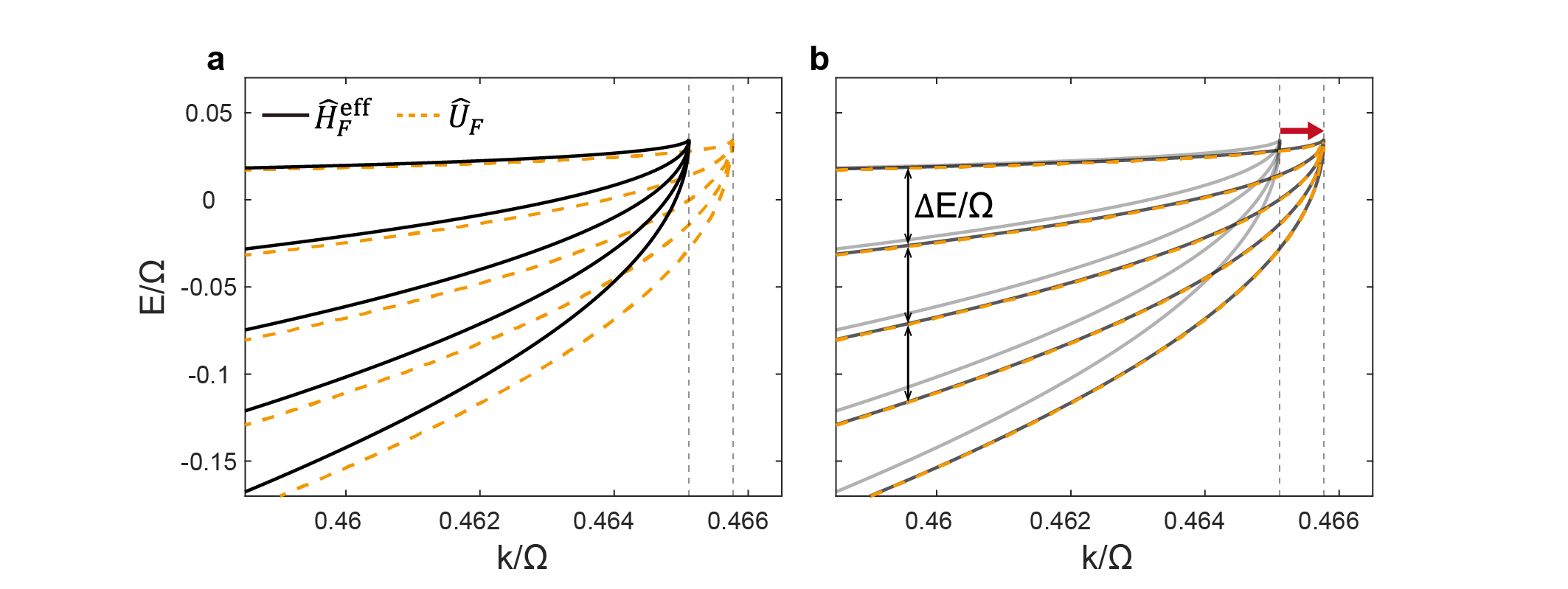}
  \caption{(a) Quasi-eigenenergy spectrum of the quantum PTC as a function of the momentum $k$. Solid lines represent the quasi-eigenenergies obtained from the effective Floquet Hamiltonian $\hat H^{\text{eff}}_F$, while dashed lines indicate the quasi-eigenenergies derived directly from the Floquet evolution operator  $\hat U_F$. The vertical dashed lines mark the momentum gap edges for two cases. (b) The spectrum calculated using $\hat U_F$ (dashed lines) and $\hat H^{\text{eff}}_F$ (solid lines) up to a momentum shift (see Appendix~\ref{appendix:D}), as highlighted by the arrows.}
  \label{fig:E_Band}
\end{figure}

\begin{figure}[htb!]
  \centering
    \includegraphics[width=1\textwidth]{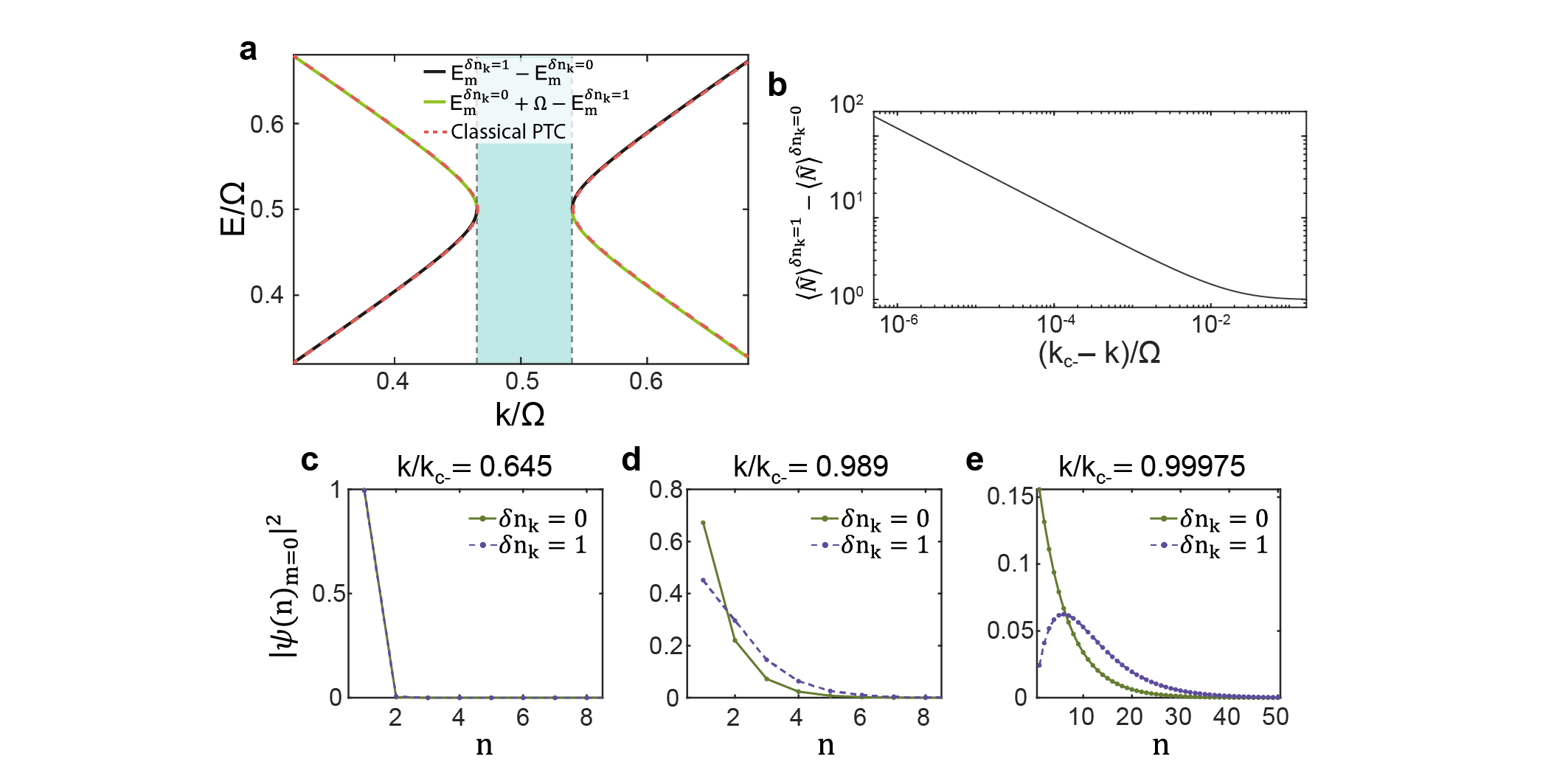}
  \caption{ 
  (a) The energy spacing between eigenstates differed by one momentum quanta precisely reproduces the spectrum of the classical PTC. (b) The difference in the expectation values of the photon number,$\langle \hat{N} \rangle$, between the eigenstates with $\delta n_k=0$ and $\delta n_k=1$. For each $m$, the eigenstate with the minimum $\langle \hat{N} \rangle$ is used ($m=0$). As $k$ approaches the edge of the momentum gap, $k_{c-}$, the difference in $\langle \hat{N} \rangle$ increases with $\langle \hat N \rangle^{\delta n_k{=}1}{-}\langle \hat N \rangle^{\delta n_k{=}0} \sim 1/|k_{c\text-}-k|^{2}$. (c-e) Probability distributions of the eigenstates $P(n)=|\prescript{}{\delta n_k}{\bra{n}}{\psi_{m=0}}\rangle|^2$ with the minimum $\langle \hat{N} \rangle$ for $\delta n_k=0$ and $\delta n_k=1$. For $k/k_{c-} \ll 1$, the probability distributions have nearly identical shapes, with a difference in $\langle \hat{N} \rangle$ of approximately 1. As $k/k_{c-}$ approaches 1, the probability distributions become increasingly distinct, resulting in a larger difference in $\langle \hat{N} \rangle$.}
  \label{fig:N_m0m1}
\end{figure}

Importantly, from the spectrum of the quantum PTC, we can reconstruct the spectrum of the classical PTC. That is, the difference between eigenenergies, $\pm(E^{\delta n_k=1}_m-E^{\delta n_k=0}_m)+l_F \Omega$ restores the whole eigenfrequency spectrum of the classical PTC. Figure~\ref{fig:N_m0m1}a demonstrates that the spectrum of the positive slope is reproduced by $ \omega_{\text{CPTC,+}}=+(E^{\delta n_k=1}_m-E^{\delta n_k=0}_m)$ and the spectrum of the negative slope is reproduced by $ \omega_{\text{CPTC,-}}= \Omega -(E^{\delta n_k=1}_m-E^{\delta n_k=0}_m)$. Other Floquet side bands can be reproduced by choosing different $l_F$.

Next, when $k$ approaches edges of the momentum gap, $k_{c-}$, the slope of the spectrum increases, accompanied by a divergence in the difference of the photon number expectation values, $\langle \hat{N} \rangle$ between $\delta n_k=0$ and $\delta n_k=1$. Figure \ref{fig:N_m0m1}b illustrates the difference in photon number expectation values, $\langle \hat{N} \rangle$, between the eigenstates for $\delta n_k=1$  and $\delta n_k=0$, along with the corresponding probability distributions of these eigenstates in Fig.~\ref{fig:N_m0m1}c-e. For each $\delta n_k$, the eigenstate with the minimum $\langle \hat{N} \rangle$ is selected ($m=0$). When $k/k_{c-} \ll 1$, the probability distributions for $\delta n_k=0$ and $\delta n_k=1$ are nearly identical (Fig.~\ref{fig:N_m0m1}c), resulting in the difference in $\langle \hat{N} \rangle$ remaining constant at approximately 1. However, as $k/k_{c-}$ approaches 1, the distributions show significant deviations, leading to a larger difference in $\langle \hat{N} \rangle$ (Fig.~\ref{fig:N_m0m1}d-e). This explains the diverging difference of the average photon number  shown in Fig.~\ref{fig:N_m0m1}b, $\langle \hat N^{\delta n_k=1}\rangle-\langle \hat N^{\delta n_k=0}\rangle$, and it indicates the amplitude of oscillation in the corresponding classical PTC diverges at the critical momentum as well. Note that this information is  available only through the analysis of the quantum PTC.

\section{\,\,\,\,\,\, Transfer matrix approach for the quantum PTC}
\label{appendix:B}

    \noindent The effective Floquet Hamiltonian $\hat{H}^{\text{eff}}_{F}$ is given by: 
    \begin{equation}
        \hat{H}^{\text{eff}}_{F} = \sum_{n=0}^{\infty} \left[(2n+\delta n_{k})k - n\Omega\right]|{n}\rangle \langle {n}| + \frac{1}{2}k\alpha_{c} \sqrt{(n+1)(n+\delta n_{k}+1)}\left[|{n}\rangle\langle {n+1}| + h.c.\right] ,
    \end{equation}
    which provides an effective description of the quantum PTC in the vicinity of the critical momentum. 
    The difficulty dealing with the Hamiltonian from a direct diagonalization is that the coefficients of $\textit{onsite}$ and $\textit{hopping}$ terms grows indefinitely as $n\rightarrow \infty$, implying that near the critical momentum and within the momentum gap the finite size effect is uncontrolled. The transfer matrix offers an alternative route by focusing on the relation of neighboring amplitudes of wave functions. 
    The recursive relation of $\psi(n)=\langle n | \psi \rangle$ can be obtained from  $\bra n \hat{H}^{\text{eff}}_{F}\ket{\psi} = E\langle n |\psi\rangle $,   \begin{equation}
        E\psi(n) = [(2n+\delta n_{k})k-n\Omega]\psi(n) + \frac{1}{2}k\alpha_{c} \sqrt{(n+1)(n+\delta n_{k} +1)}\psi(n+1) + \frac{1}{2}k\alpha_{c} \sqrt{n(n+\delta n_{k})}\psi(n-1).
    \end{equation}
    The transfer matrix connects  the neighboring amplitudes of wave function: 
\begin{equation}
\begin{bmatrix}
    \psi(n+1) \\
    \psi(n)
\end{bmatrix}
=
T_{n}
\begin{bmatrix}
    \psi(n) \\
    \psi(n-1)
\end{bmatrix}
=T_{n}\Psi(n-1),
\end{equation}
where the transfer matrix $T_n$ is
\begin{equation}
T_{n} =
\begin{bmatrix}
    \frac{2}{\alpha_{c} k} \frac{\left( E + n(\Omega - 2k) - k\delta n_{k} \right)}{(n+1)\sqrt{1 + \delta n_{k}/(n+1)}} & -\frac{n\sqrt{1 + \delta n_{k}/n}}{(n+1)\sqrt{1 + \delta n_{k}/(n+1)}} \\
    1 & 0
\end{bmatrix}.
\end{equation}
For a given momentum $k$ and energy $E$, the transfer matrix allows us to generate $\psi(n)$ from $\psi(n=0)$. As a result, the method is critically used for the accurate estimation of the Lyapunov exponent near the quantum phase transition to characterize  universality classes~\cite{Ishii1973Localization, mackinnon1983scaling, J.L.Pichard_1981Finitesizescaling, crit2009slev,Dwivedi2016generalized, Luo2021Transfermatrix,Zhang2022Lyapunovexponent, anis2023xiao}. 
The transfer matrix, while initially appearing complex, undergoes significant simplification in the limit as $n \to \infty$. The asymptotic form of the transfer matrix is:
\begin{equation}
\lim_{n \to \infty} T_{n} = T =
\begin{bmatrix}
    \frac{2}{\alpha_{c}} \left(\frac{\Omega}{k} - 2\right) & -1 \\
    1 & 0
\end{bmatrix}
=
\begin{bmatrix}
    P & -1 \\
    1 & 0
\end{bmatrix},
\label{eq:converged_transfermatrix}
\end{equation}
where $Z = \frac{2}{\alpha_{c}} \left(\frac{\Omega}{k} - 2\right)$. This expression highlights that, in the limit as $n \to \infty$, the transfer matrix  becomes independent of both the energy and total momentum $P=\delta n_k$. The physical reason of taking this limit $n\rightarrow \infty$ is from the anticipation that when the eigenstates of the quantum PTC is delocalized in the Floquet-photonic space, the quantum dynamics taking place at larger $n$ region is more important and eventually dominant as the strength of hopping is greater. This expectation is confirmed by matching our analytical prediction and numerical results.

\noindent In the classical PTC, the in-gap states lead to significant light amplification, while in the quantum PTC, this amplification manifests as the delocalization of the photon wave function across the Floquet-photonic lattice. We employ the transfer matrix to rigorously define the exact critical momentum at which the localization-delocalization transition occurs in the quantum PTC, thereby introducing the concept and demonstrating the analytical method for studying this transition. 

\noindent The Lyapunov exponent in terms of the transfer matrix is defined as:
    \begin{equation}
        \gamma = \lim_{N \rightarrow \infty} \frac{1}{N} \ln\left[\frac{|\Psi(N)|}{|\Psi(0)|}\right] = \lim_{N \rightarrow \infty} \frac{1}{2N} \ln \left[ \frac{\Psi^{\dg}(0) T_{1}^{\dagger}T_{2}^{\dagger}T_{3}^{\dagger}\cdots T_{N}^{\dagger} T_{N}\cdots T_{3}T_{2}T_{1} \Psi(0)}{\Psi^{\dagger}(0)\Psi(0)}\right].
    \end{equation}
where $|\Psi(n)| = \sqrt{|\psi(n+1)|^2 + |\psi(n)|^2}$. If the wave function is localized, the Lyapunov exponent $\gamma$ is nonzero and positive; in contrast, a zero Lyapunov exponent signifies delocalization, defining the region known as the metallic phase. 
 To simplify the computation, the transfer matrix sequence can be truncated at an intermediate point $q$, where $0<q<N$ and $\Psi^{\dg}(q)\Psi{(q)} \neq 0 , \infty$. In this case, the Lyapunov exponent is equivalently expressed as:
\begin{align}
    \gamma &= \lim_{N\rightarrow \infty} \frac{1}{2N} \ln\left[\frac{\Psi^{\dg}(q)T_{q+1}^{\dagger}T_{q+2}^{\dagger}\cdots T_{N}^{\dagger}T_{N}\cdots T_{q+2}T_{q+1}\Psi(q) }{\Psi^{\dg}(q)\Psi(q)} \frac{\Psi^{\dg}(q)\Psi(q)}{\Psi^{\dg}(0)\Psi(0)}\right] \notag \\
    &= \lim_{N\rightarrow \infty} \frac{1}{2N} \ln\left[\frac{\Psi^{\dg}(q)T_{q+1}^{\dagger}T_{q+2}^{\dagger}\cdots T_{N}^{\dagger}T_{N}\cdots T_{q+2}T_{q+1}\Psi(q) }{\Psi^{\dg}(q)\Psi(q)} \right],
\end{align}
    where $\Psi(q) = T_{q}T_{q-1}T_{q-2}\cdots T_{1}\Psi(0)$. Under the condition that $\Psi^{\dg}(q)\Psi(q) \neq 0, \infty$, the term $\left[\Psi^{\dg}(q)\Psi(q)/\Psi^{\dg}(0)\Psi(0)\right]$ can be factored out as an additive constant outside the logarithm. As $N \to \infty$, $q$ remains finite, causing the separated term to vanish in the limit and leaving the Lyapunov exponent unaffected. Thus, the starting point of the calculation does not affect the determination of the critical point or the critical exponent. Now, we select $q$ to be sufficiently large so that the transfer matrix converges to its asymptotic form, as given in Eq.~\eqref{eq:converged_transfermatrix}. This allows us to express the Lyapunov exponent as  
\begin{equation}
    \gamma \approx \lim_{N \rightarrow \infty} \frac{1}{2N} \ln \left[ \frac{\Psi^{\dg}(q)(T^{\dagger}T)^{N-q}\Psi(q)}{\Psi^{\dg}(q)\Psi(q)} \right].
\end{equation}
The Lyapunov exponent can be expressed solely using Eq.~\eqref{eq:converged_transfermatrix}, indicating that the critical point and the critical exponent depend only on the momentum $k$ and the modulation parameters $\alpha_c, \Omega$. The eigenvalue equation of the transfer matrix is given by:
\begin{align}
        T\phi_{\pm} = \lambda_{\pm}\phi_{\pm}, \qquad
        \lambda_{\pm} = \frac{Z \pm \sqrt{Z^{2} - 4}}{2}.
    \end{align}
    Due to probability conservation, the transfer matrix is a symplectic matrix, meaning its eigenvalues satisfy the condition: $\lambda_{+}\lambda_{-}=1$. 
    The $q$-th wave function can be decomposed into the eigenbasis of the transfer matrix as $\Psi(q) = C_{+}\phi_{+} + C_{-}\phi_{-}$ where $\phi_\pm$ are not orthogonal and $C_{\pm}\neq 0$, in general. By plugging in, the Lyapunov exponent is given by:
    \begin{equation}
        \gamma = \lim_{N \rightarrow \infty} \frac{1}{2N} \ln \left[ \frac{|\lambda_{+}^{N-q}C_{+}|^{2} + (\lambda_{+}^{*}\lambda_{-})^{N-q}C_{+}^{*}C_{-}\phi_{+}^{\dg}\phi_{-} + (\lambda_{+}\lambda_{-}^{*})^{N-q}C_{+}C_{-}^{*}\phi_{+}\phi_{-}^{\dg} + |\lambda_{-}^{N-q}C_{-}|^{2}}{|C_{+}|^{2} + C_{+}^{*}C_{-}\phi_{+}^{\dg}\phi_{-} + C_{+}C_{-}^{*}\phi_{+}\phi_{-}^{\dg}+|C_{-}|^{2}} \right].
        \label{eq:lyapunov_calculation}
    \end{equation}
    When $Z^2 > 4$, the eigenvalues $\lambda_{\pm}$ are both real and separated by a square-root term. To capture the dominant contribution, we define $\lambda \equiv \max\bigl(\lvert\lambda_{+}\rvert,\,\lvert\lambda_{-}\rvert\bigr)$. The Lyapunov exponent $\gamma$ is then governed by $\lambda$ because, as $N \to \infty$, $\lambda^{2N}$ dominates the exponential growth of the wave function norm in the Floquet-photonic space. Hence, we find that $\gamma = \ln(\lambda)$, which quantifies the rate of exponential localization.  On the other hand, when $Z^2 \le 4$, the eigenvalues lie on the unit circle (pure phase rotations), so $\gamma = 0$. To summarize, the Lyapunov exponent is    
  \[\gamma= 
   \begin{cases}
   0, & Z^{2} \leq 4,\\
   \ln\lambda =  \ln \left[ |\frac{Z}{2}| + \sqrt{\left(\frac{Z}{2}\right)^2 - 1}\right] = \text{arcosh}(|\frac{Z}{2}|), & Z^{2} > 4.
   \end{cases}\]
   Because $|\psi(n)| \simeq e^{-\gamma n}|\psi(0)|$ for $n\gg 1$, $\gamma=0$ corresponds to a delocalized phase, while $\gamma>0$ corresponds to a localized phase. The phase transition between the two occurs at the critical momentum, and it is determined by the condition $Z^{2} = 4$. Solving this yields
   \begin{equation}
       \quad k_{c-} = \frac{\Omega}{2 + \alpha_{c}}, \quad k_{c+} = \frac{\Omega}{2 - \alpha_{c}}. 
   \end{equation}
   Meanwhile, the localization length $\xi$ is defined as the inverse of the Lyapunov exponent $\gamma$. Specifically,    \begin{equation}
        \xi^{-1}= \gamma= 
       \text{arcosh} \left(\left|\frac{1-\Omega/2k}{1-\Omega/2k_{c\pm}}\right|\right), \quad k < k_{c-} \; \text{or} \; k> k_{c+}.
       \label{eq:localization_length_function}
   \end{equation}
Equation~\eqref{eq:localization_length_function} expresses the localization length as a function of momentum. It shows that $\xi$ diverges as $k$ approaches $k_{c-}$ from below $(k \to k_{c-}+0^{-})$ and also as $k$ approaches $k_{c+}$ from above $(k \to k_{c+}+0^{+})$. The following equation defines the critical exponent $\nu$, which characterizes how the localization length $\xi$ diverges near the critical point.
    \begin{equation}
       \nu = -\lim_{\delta \rightarrow 0} \frac{\ln\xi}{\ln\delta}, \quad \delta \equiv \left| \frac{k_{c\pm}-k}{k_{c\pm}} \right|.
       \label{eq:def_critical_exponent}
    \end{equation}
    To analyze the behavior of $\xi$ near $\lambda = 1$ (or $\gamma=0$), we consider a Taylor expansion of $x\xi$ in terms of $x \equiv \lambda - 1$. Dividing out the leading factor $x$ yields
    \begin{equation}
        \xi = \frac{1}{\ln\lambda} = \frac{1}{\ln(1+x)} = \frac{1}{x} + \frac{1}{2} + \dots, \quad (x \equiv \lambda -1).
    \end{equation}
    Now, focus on the regime $k < k_{c-}$ near the critical point. By examining how $x$ scales with $\delta \equiv \bigl|\tfrac{k_{c-} - k}{k_{c-}}\bigr|$, one obtains
    \begin{align}
        x = \delta \left[ C(\Omega,k_{c-}) \frac{1}{1-\delta} +  \frac{1}{\sqrt{\delta}}\sqrt{C(\Omega,k_{c-})\frac{1}{1-\delta}\left(\frac{Z}{2}+1\right)} \right],
    \end{align}
     where $C(\Omega,k_{c-})$ is a non-singular function. Because the highest power of $\delta$ in this expression is $\delta^1$, the critical exponent $\nu$ follows as
     \begin{equation}
        \nu = -\lim_{\delta \rightarrow 0} \frac{\ln\xi}{\ln\delta} = \lim_{\delta \rightarrow 0} \frac{\ln x}{\ln\delta} = 1.
    \end{equation}
\newline
    \noindent 
    Next, we establish the exponential growth of photonic energy within the momentum gap. The \textit{classical} field growth rate $\gamma_{\text{cl}}$ within the momentum gap of PTCs can be calculated from the imaginary part of the eigenfrequencies. Correspondingly, the photonic energy growth rate is $2\gamma_{\text{cl}}$. Here we  proceed to derive the quantum counterpart of this result. In the analysis above, we focused on the localized regime as $k$ approaches the critical point. At $k = k_{c\pm}$, the eigenvalues of the transfer matrix coalesce at $\lambda_{\pm} = 1$. Once the momentum enters the gap ($k \in \mathcal{K}_{\text{gap}}$), the eigenvalues can be written as $\lambda = e^{\mp i\phi}$, where $\phi = \pm \arccos\bigl(Z/2\bigr)$ and $Z = e^{i\phi} + e^{-i\phi}$. Under these conditions, the asymptotic transfer matrix from Eq.~\eqref{eq:converged_transfermatrix} admits two eigenvectors, $\frac{1}{\sqrt 2}\left[e^{i\phi} \;\, 1\right]^{T}$ and $\frac{1}{\sqrt 2}\left[{e^{-i\phi} \;\, 1}\right]^{T}$. Each state satisfies $\psi_t(n+1) = e^{\pm i\phi}\,\psi_t(n)$, indicating a freely propagating wave whose phase rotates uniformly from site $n$ to site $n+1$. This phase progression sustains a net current within the Floquet--photonic lattice, ultimately causing the photonic energy to grow continuously. We define the photonic energy operator in the photon-number basis as $\hat{H}_{\text{ph}} = \text{diag}(0,2k,4k,\cdots) = k\hat{N}$, under the assumption $\delta n_{k}=0$. Its expectation value evolves according to $ d\braket{\hat{H}_{\text{ph}}}/dt=-i\bra{\psi_{t}}\left[\hat{H}^{\text{eff}}_{F},\hat{H}_{\text{ph}}\right]\ket{\psi_{t}} =
    \Sigma_{n} 2(n+1)\alpha_{c} k^2 \Im\left[\psi_{t}^{*}(n)\psi_{t}(n+1)\right]$. This shows that phase rotations across the photon‐number basis directly produce the photonic energy current. By analogy with classical PTCs, the quantum exponential growth rate emerges as
    \begin{align}
        \label{eq:exp_growing_rate}
        \frac{1}{\braket{\hat{H}_{\text{ph}}}}\frac{d\braket{\hat{H}_{\text{ph}}}}{dt} 
        &= \frac{\sum_{n}2(n+1)\alpha_{c} k^2 \Im\left[\psi_{t}^{*}(n)\psi_{t}(n+1)\right]}{\sum_{n}2nk\left[\psi_{t}^{*}(n)\psi_{t}(n)\right]}
        \simeq \pm \frac{\sum_{n}2(n+1)\alpha_{c} k^2 \sin\phi\left[\psi_{t}^{*}(n)\psi_{t}(n)\right]}{\sum_{n}2nk\left[\psi_{t}^{*}(n)\psi_{t}(n)\right]} \simeq \pm\alpha_{c} k \sin\phi,
    \end{align}
    which establishes the quantum analog of exponential photonic‐energy growth observed in classical PTCs. Expressing $\phi$ in terms of momentum yields: 
    \begin{equation}
        \gamma_{\text{qu}} = \pm \alpha_{c} k \sin\phi = \pm \alpha_{c} k \sqrt{1 - \left(\frac{1-\Omega/2k}{1-\Omega/2k_{c-}}\right)^2} = \pm2k\sqrt{\left(\frac{\Omega}{2k_{c-}}-1\right)^2 - \left(\frac{\Omega}{2k} - 1\right)^2}.
    \end{equation}
\begin{figure}[htb!]
  \centering
  \includegraphics[width=0.9\textwidth]{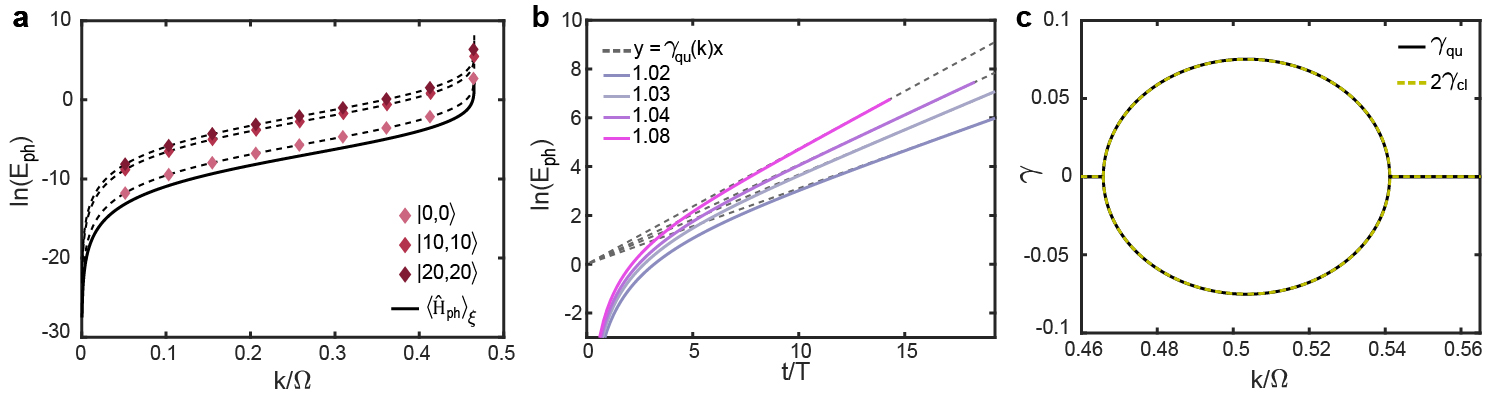}
  \caption{\label{fig:exponent}(a) Relation between the localization length and photonic energy oscillation amplitude in localized regime $k \in \mathcal{K}_{\text{band}}$, Fig.~\ref{fig3}b. The states in the legend means the initial position of the wave function in 1-dimensional Floquet-photonic lattice $\{\ket{n,n;l_{F}=n}\}$ with $\delta n_{k} = 0$. $\braket{\hat{H}_{\text{ph}}}_{\xi}$ is photonic energy expectation value from the ansatz $|\psi_{\xi}(n)| \sim e^{-n/\xi}$. The data points (diamond marks) are the difference between maximum value and minimum value of the photonic energy oscillation, which is twice of oscillation amplitude. (b) Relation between the $\gamma_{\text{qu}}$ and exponential growing rate in Fig.~\ref{fig3}c, in $k \in \mathcal{K}_{\text{gap}}$. The colored lines represent the $\ln({E}_{\text{ph}}(t))$ shifted along the y-axis at momentum indicated in the legend: $k = 1.02k_{c-}, 1.03k_{c-}, 1.04k_{c-}, 1.08k_{c-}$. An initial ($t=0$) quantum state is placed at $\ket{0,0;l_{F}=0}$. (c) Relation between the $\gamma_{\text{qu}}$ and $2\gamma_{\text{cl}}$. The $\gamma_{\text{qu}}$ is shifted along the x-axis due to the overall momentum shift between the classical PTC and $\hat{H}^{\text{eff}}_{F}$ (Appendix \ref{appendix:A}, \ref{appendix:D}). The numerical simulations are conducted with the condition of $\alpha_{c} = 0.15$.}
  \label{fig:B1}
\end{figure}

    In the localized regime, a wave packet initially placed at $\ket{n,n;\,l_F = n}$ undergoes oscillatory motion in the Floquet--photonic space, driven purely by phase interference among the $\hat{H}^{\text{eff}}_F$ eigenstates (see Fig.~\ref{fig3}b). The oscillation frequency of the photonic energy matches the uniform energy spacing of the quasi-eigenenergy spectrum (see Appendix~\ref{appendix:F} for details). Moreover, the oscillation amplitude is directly tied to the wave‐packet spread in the Floquet–photonic space, which is quantified by the localization length. By adopting the ansatz $|\psi_{\xi}(n)| = 
    A e^{-n/\xi}$ where $A=\sqrt{1-e^{-2/\xi}}$, 
    the photonic energy expectation value $\braket{\hat{H}_{\text{ph}}}_{\xi} = \bra{\psi_{\xi}}\hat{H}_{\text{ph}}\ket{\psi_{\xi}}$, is proportional to the oscillation amplitude as shown in  Fig.~\ref{fig:B1}a. In the log-scale plot, the oscillation amplitude parallels $\langle \hat{H}_{\text{ph}}\rangle_{\xi}$ up to a constant vertical shift, demonstrated across three distinct initial wave-packet positions.
    
In the momentum-gap region $k \in \mathcal{K}_{\text{gap}}$, the wave function delocalizes over the Floquet--photonic basis, causing the photonic energy to grow exponentially (see Fig.~\ref{fig3}c). The growth rate of this energy directly reflects the phase-rotation rate (Eq.~\eqref{eq:exp_growing_rate}).
    Figure~\ref{fig:B1}b demonstrates excellent agreement between our analytic prediction and the numerical results obtained by evolving a wave packet initially at $\ket{0,0;\,l_F=0}$. Equation~\eqref{eq:exp_growing_rate} remains valid for $n \gg 1$, and the plot shows strong convergence of the numerical data at later times $(t/T > 10)$.    
    Lastly, Fig.~\ref{fig:B1}c demonstrates that $\gamma_{\text{qu}}$ which we analytically computed above  makes an excellent matching with the imaginary part of eigenfrequencies $\gamma_{\text{cl}}$ with the extra factor of two. It is because the energy of electromagnetic waves in the classical PTC is $E_{\text{ph}}^{\text{classic}}(t)  \sim e^{2\gamma_{\text{cl}}t}$.  Without any introduction of non-Hermitian components, the behavior of exponentially growing field strength within the momentum gap is explained in the quantum PTC by computing the photonic energy current when eigenstates are delocalized in the Floquet-photonic space.

\section{\,\,\,\,\,\, Transfer matrix approach for the classical PTC}
\label{appendix:C}
    \noindent In a PTC, Maxwell's equations dictate how the electromagnetic fields respond to time-varying permittivity. When the permittivity $\varepsilon(t)$ varies with time, these equations take the form
\begin{equation}
\nabla \times \bold{E}
\,=\,
-\,\frac{\partial \bold{B}}{\partial t},
\quad
\nabla \times \bold{B}
\,=\,
\frac{\partial}{\partial t}\bigl[\varepsilon(t)\,\bold{E}\bigr],
\end{equation}
thereby illustrating the direct influence of temporal modulations in $\varepsilon$ on $\bold{E}$ and $\bold{B}$. For simplicity, we assume a plane wave traveling along the $x$ direction. In this configuration, the electric field oscillates along the $y$ axis, while the magnetic field oscillates along the $z$ axis. We then obtain a differential equation for the out-of-plane magnetic field $B_z$ \cite{Wang2018photonic}, which can be recast in a Schr\"odinger-like form:
\begin{equation}
k^{2} B_{z} \;=\; \frac{\partial}{\partial t}\!\Bigl[\varepsilon\,\frac{\partial B_{z}}{\partial t}\Bigr],
\qquad
i\,\frac{\partial}{\partial t}
\begin{bmatrix}
B_{z}\\[6pt]
\varepsilon\,\dot{B}_{z}
\end{bmatrix}
\;=\;
\begin{bmatrix}
0 & i\,\varepsilon^{-1}\\[4pt]
-\,i\,k^{2} & 0
\end{bmatrix}
\begin{bmatrix}
B_{z}\\[4pt]
\varepsilon\,\dot{B}_{z}
\end{bmatrix}
\;=\;
\hat{H}_{\text{eff}}
\begin{bmatrix}
B_{z}\\[3pt]
\varepsilon\,\dot{B}_{z}
\end{bmatrix}.
\end{equation}
The resulting effective Hamiltonian is both non-Hermitian and time periodic, making it amenable to Floquet theory. In particular, we apply the transform indicated in Eq.~(\ref{eq:floquettransform}) to recast the system into a higher-dimensional representation. We assume a single-frequency modulation of the inverse permittivity, $\varepsilon^{-1}(t) = 1 - 2\alpha_{c}\cos(\Omega t).$ This periodic function $\varepsilon(t)$ induces a Floquet Hamiltonian of the form
\begin{align*}
\Tilde{H}_{F}(k)
&=\sum_{n=-\infty}^{\infty}
\begin{pmatrix}
    -n\Omega & i \\
    -ik^{2} & -n\Omega
\end{pmatrix}
|{n}\rangle \langle {n}|
+
\begin{pmatrix}
    0 & -\,i\,\alpha_{c} \\
    0 & 0
\end{pmatrix}
|{n}\rangle \langle {n+1}|
+
\begin{pmatrix}
    0 & -\,i\,\alpha_{c} \\
    0 & 0
\end{pmatrix}
|{n+1}\rangle \langle {n}|,
\nonumber\\
&=\sum_{n=-\infty}^{\infty} M_{n}\,|{n}\rangle \langle {n}|
\;+\;
J\,|{n}\rangle \langle {n+1}|
\;+\;
J\,|{n+1}\rangle \langle {n}|.
\end{align*}
Note that $n$ is Floquet index only in contrast to the case of the quantum PTC where the photon number state $\ket{n_k,n_{-k}}$ is present in addition to the Floquet index $l_F$. 
The hopping matrix $J$ is non-invertible and nilpotent, satisfying $J^2 = 0$ and $\det(J) = 0$. Consequently, the generalized transfer matrix method \cite{Dwivedi2016generalized} needs to be employed. We choose the singular value decomposition of $J$ as $J = V\Xi W^{\dagger} = \left[0 \, -i\right]^{T}\alpha\left[0 \;\, 1\right]$, where $V$ and $W$ span $\mathbb{C}^2$ as an orthonormal basis. From this construction, one obtains the following transfer-matrix relation:
\begin{equation}
        \begin{bmatrix}
            W^{\dagger} \psi(n+1) \\
            W^{\dagger} \psi(n)
        \end{bmatrix}
        =
        T_{n}
        \begin{bmatrix}
            W^{\dagger} \psi(n) \\
            W^{\dagger} \psi(n-1)
        \end{bmatrix}
        =
        \begin{bmatrix}
            \frac{1}{\alpha} \left[ 1 - \left( \frac {E - n\Omega}{k}\right)^{2} \right] & -1 \\
            1 & 0
        \end{bmatrix}
        \begin{bmatrix}
            \varepsilon \dot{B_{z}}(n) \\
            \varepsilon \dot{B_{z}}(n-1)
        \end{bmatrix}.
    \end{equation}
    By diagonalizing, one can verify that the eigenvalues of $\tilde H_F$ are real and distinct within the band $k\in\mathcal K_{\text{band}}$. Inside the momentum gap $k\in \mathcal K_{\text{gap}}$, the eigenvalues remain complex-conjugate pairs, each acquiring an imaginary part that induces exponential growth or decay. At the momentum gap edge, these eigenvalues coalesce into a single value, marking an exceptional point of classical PTCs. At the gap edge, two distinct Floquet states merge into a single wave function for each Floquet index, rendering all eigenstates at this exceptional point symmetric across the Floquet indices. This symmetry implies that the transfer matrix acquires a unit eigenvalue when multiplied over a symmetric pair.
    For instance, consider the momentum gap around $E/\Omega = \tfrac12$. Here, the product $T_{1} T_{0}$ couples the states 
$W^\dagger\psi(0),\,W^\dagger\psi(-1)$ with $W^\dagger\psi(2),\,W^\dagger\psi(1)$. That is,
\begin{equation}
        \begin{bmatrix}
            W^{\dagger} \psi(2) \\
            W^{\dagger} \psi(1)
        \end{bmatrix}
        =
        T_{1}T_{0}
        \begin{bmatrix}
            W^{\dagger} \psi(0) \\
            W^{\dagger} \psi(-1)
        \end{bmatrix}. 
\end{equation}        
At the critical momentum, $\psi(2)$ and $\psi(-1)$, as well as $\psi(1)$ and $\psi(0)$, share the same norm. Owing to this symmetric structure, the wave number associated with the unity eigenvalue denotes the transition. Extending the approach by computing $T_2T_1T_0T_{-1}$ yields a more precise estimate of the critical point. In this case, the analytic expression for the first-order momentum-gap edge is 
    \begin{equation}
        k_{c-} \approx \frac{\Omega}{2 + \alpha_{c} - \frac{1}{8}\alpha_{c}^{2}-\frac{5}{64}\alpha_{c}^{3}},
        \label{eq:SM_classical_critical}
    \end{equation}
and can be further improved by including higher-order products of the transfer matrix.
    
    The position of the momentum gap edge in a classical PTC described by $\varepsilon = \varepsilon_0 + \varepsilon_r \cos(\Omega t)$ has been analyzed in Ref.~\cite{Wang2018photonic}. In the regime where $\bigl|\varepsilon_r/\varepsilon_0\bigr|\ll 1$ and $\varepsilon_0 = 1$, one obtains the approximate expression $k_{-} = \Omega(2-\varepsilon_{r})/(4-\varepsilon_{r})$. Ignoring the second order and higher in the analysis, $\tilde H_{n\ge 2} =0$, the $k_-$ is actually a perturbative estimation for $\varepsilon(t)=\varepsilon_0/(1-\frac{\varepsilon_r}{\varepsilon_0} \cos\Omega t)$.     
    To compare with our result from the transfer matrix, let us Taylor expand the inverse of $k_-$: 
    \begin{align}
        k_-^{-1} = \frac{1}{\Omega} \left(\frac{4-\epsilon_r}{2-\epsilon_r}\right) \simeq \frac{1}{\Omega} \left(2+ \frac{\epsilon_r}{2} +\frac{\epsilon_r^2}{4}+O(\epsilon_r^3) \right),
    \end{align}
    which is consistent with $k_{c-}^{-1}$ up to the first order  when we identify $\alpha_c = \epsilon_r/2$, but the second order correction deviates.  Our derivation of Eq.~(\ref{eq:SM_classical_critical}) relies on the symmetric structure of eigenvectors in the Floquet space and it provides more accurate results. The precision can be further enhanced by multiplying more transfer matrices, see Appendix~\ref{appendix:D}.

To extend the analysis, one can introduce a second harmonic in the inverse permittivity, $\varepsilon^{-1} = 1 - 2\alpha_{c}\cos(\Omega t) - 2\beta_{c}\cos(2\Omega t)$, where $\beta_{c}$ generates an additional hopping term in the Floquet Hamiltonian. Building on this extension, the Floquet Hamiltonian takes the form
\begin{align}
       \Tilde{H}_{F}(k)
        &=\sum_{n=-\infty}^{\infty}
        \begin{pmatrix}
            -n\Omega & i \\
            -ik^{2} & -n\Omega
        \end{pmatrix}
        |{n}\rangle \langle {n}|
        +
        \begin{pmatrix}
            0 & -i\alpha_{c} \\
            0 & 0
        \end{pmatrix}
        |{n}\rangle \langle {n+1}|
        +
        \begin{pmatrix}
            0 & -i\alpha_{c} \\
            0 & 0
        \end{pmatrix}
        |{n+1}\rangle \langle {n}|\\\nonumber
        & \qquad \qquad
        +
        \begin{pmatrix}
            0 & -i\beta_{c} \\
            0 & 0
        \end{pmatrix}
        |{n}\rangle \langle {n+2}|
        +
        \begin{pmatrix}
            0 & -i\beta_{c} \\
            0 & 0
        \end{pmatrix}
        |{n+2}\rangle \langle {n}|,\\        
        &
        =
        \sum_{n'=-\infty}^{\infty}
        \begin{pmatrix}
            -n\Omega & i & 0 & -i\alpha_{c} \\
            -ik^2 & -n\Omega & 0 & 0 \\
            0 & 0 & -(n+1)\Omega & i \\
            -i\alpha_{c} & 0 & -ik^2 & -(n+1)\Omega 
        \end{pmatrix}
        |{n'}\rangle \langle {n'}|
        +
        \begin{pmatrix}
            0 & -i\beta_{c} & 0 & 0 \\
            0 & 0 & 0 & 0 \\
            0& -i\alpha_{c} & 0 & -i\beta_{c} \\
            0 & 0 & 0 & 0
        \end{pmatrix}
        |{n'}\rangle \langle {n'+1}|& \\\nonumber
        & \qquad \qquad
        +
        \begin{pmatrix}
            0 & -i\beta_{c} & 0 & -i\alpha_{c} \\
            0 & 0 & 0 & 0 \\
            0& 0 & 0 & -i\beta_{c} \\
            0 & 0 & 0 & 0
        \end{pmatrix}
        |{n'+1}\rangle \langle {n'}|,\\        
        &
        =\sum_{n'=-\infty}^{\infty}
        M_{n'}|{n'}\rangle \langle {n'}| + J_{p}|{n'}\rangle \langle {n'+1}| + J_{m}|{n'+1}\rangle \langle {n'}|,
    \end{align}
    where the enlarged unit cell is introduced that $\ket{n'}$ includes both $\ket n$ and $\ket{n+1}$. 
    By extending the basis to include both the original set of states and an offset copy, we obtain a block-structured Hamiltonian that can be treated with the transfer matrix method. The resulting hopping matrices $J_{p}$ and $J_{m}$ are nilpotent, with $\det(J_{p})=\det(J_{m})=0$ and $J_{p}J_{p}=J_{m}J_{m}=0$. Consequently, their singular value decompositions, $J_{p} = V_{p} \Xi W_{p}^{\dagger}, \;$ $J_{m} = V_{m} \Xi W_{m}^{\dagger}$, satisfy  $V_{p}^{\dagger}W_{p} = V_{m}^{\dagger}W_{m} = 0$ and $V_{p}^{\dagger}V_{p} = V_{m}^{\dagger}V_{m} = W_{p}^{\dagger}W_{p} = W_{m}^{\dagger}W_{m} = \mathds{1}$.  
    Defining
    \begin{align}
        &\lambda_{\pm} = \beta_{c}^{2} + \frac{\alpha_{c}^{2}}{2} + \sqrt{ \alpha_{c}^{2} \beta_{c}^{2} + \frac{\alpha_{c}^{4}}{4}},
        & f = \frac{\lambda_{+} - \beta_{c}^{2}}{\alpha_{c}\beta_{c}} = \frac{\alpha_{c}}{2\beta_{c}} + \sqrt{1 + \frac{\alpha_{c}^{2}}{4\beta_{c}^{2}}},
    \end{align}
one obtains $\Xi = \text{diag} (\sqrt{\lambda_{+}},\sqrt{\lambda_{-}})$ and 
    \begin{equation}
        V_{p} = \frac{1}{\sqrt{1 + f^2}} 
        \begin{pmatrix}
            1 & -f \\
            0 & 0 \\
            f & 1 \\
            0 & 0
        \end{pmatrix},\,
        W_{p} = \frac{1}{\sqrt{1+f^2}}
        \begin{pmatrix}
            0 & 0 \\
            if & -i \\
            0 & 0 \\
            i & if
        \end{pmatrix},\,
        V_{m} = \frac{1}{\sqrt{1+f^2}}
        \begin{pmatrix}
            f & -1 \\
            0 & 0 \\
            1 & f \\
            0 & 0
        \end{pmatrix},\,
        W_{m} = \frac{1}{\sqrt{1+f^2}}
        \begin{pmatrix}
            0 & 0 \\
            i & -if \\
            0 & 0 \\
            if & i
        \end{pmatrix}.
    \end{equation}
The onsite green function $G = (E\mathds{1} - M_{n'})^{-1}$ becomes
    \begin{align}
        &(\det(M_{n+1})\det(M_{n}) - \alpha_{c}^2 k^4)G \, =\\\nonumber
        &\begin{pmatrix}
            [E+n\Omega]\det(M_{n+1}) & i[\det(M_{n+1}) + \alpha_{c}^2 k^2] & -\alpha_{c}k^2 [E+n\Omega] & -i\alpha_{c} [E+n\Omega] [E+(n+1)\Omega] \\
            -ik^2\det(M_{n+1}) & [E+n\Omega]\det(M_{n+1}) & i\alpha_{c}k^4 & -\alpha_{c}k[E+(n+1)\Omega]\\
            -\alpha_{c} k^2 [E+(n+1)\Omega] & -i\alpha_{c}[E+n\Omega][E+(n+1)\Omega] & [E+(n+1)\Omega]\det(M_{n}) & i[\det(M_{n}) + \alpha_{c}^2 k^2]\\
            i\alpha_{c}k^4 & -\alpha_{c} k^2 [E+n\Omega] & -ik^2\det(M_{n}) & [E+(n+1)\Omega]\det(M_{n})
        \end{pmatrix}.
        &&
    \end{align}
Collecting these elements yields the transfer matrix
\begin{equation}
        T_{n'} = 
        \begin{bmatrix}
            (W_{p}^{\dagger}GV_{P}\Xi)^{-1} & -(W_{p}^{\dagger}GV_{P}\Xi)^{-1}(W_{p}GV_{m}\Xi)\\
            (W_{m}^{\dagger}GV_{p}\Xi)(W_{p}^{\dagger}GV_{P}\Xi)^{-1} & (W_{m}^{\dagger}GV_{m}) - (W_{m}^{\dagger}GV_{p}\Xi)(W_{p}^{\dagger}GV_{P}\Xi)^{-1}(W_{p}GV_{m}\Xi)
        \end{bmatrix}.
    \end{equation}
By multiplying symmetric sets of this transfer matrix, one can obtain an exact analytical expression for the critical point to the desired precision.

\section{\,\,\,\,\,\, Momentum shift between classical and effective Floquet Hamiltonian approaches}
\label{appendix:D}
\begin{figure}[htb!]
  \centering
    \includegraphics[width=0.35\textwidth]{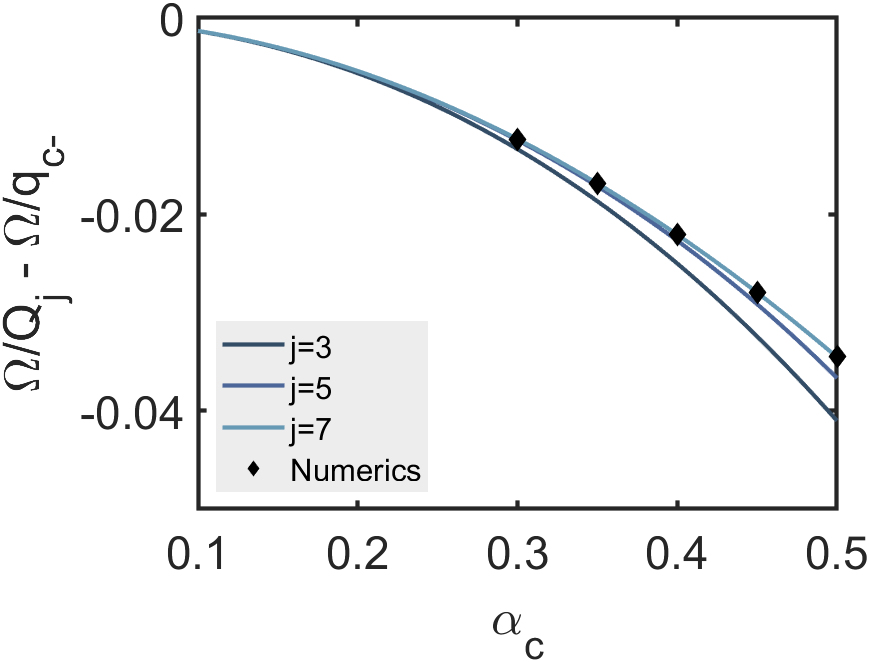}
  \caption{\label{fig:momentum_shift} Quantifying the critical-momentum shift. The symbols $Q_{c-}$ and $q_{c-}$ represent analytic expressions for the critical momentum derived from the classical PTC and $\hat{H}^{\text{eff}}_F,$ respectively. The subscript $j$ denotes the highest order of $\alpha_{c}$ considered. Black diamond points show the numerical difference between the inverse critical points from $\hat{U}_F$ and $\hat{H}^{\text{eff}}_F$.}
  \label{fig:D1}
\end{figure}

    \noindent The critical momentum  computed from the Floquet operator $\hat{U}_F$ and the classical PTC Hamiltonian $\tilde H_F$ (with higher-order terms included) coincide, while there is an overall momentum shift when using the effective Floquet Hamiltonian $\hat{H}^{\text{eff}}_F$. This shift occurs because Floquet-photonic sites with onsite energy away from $E=(\delta n_k) \Omega/2$ at $k=\Omega/2$ are intentionally dropped in the effective model (Appendix~\ref{appendix:A}). Specifically, using the transfer matrix method, the critical momentum computed from $\hat{H}^{\text{eff}}_F$ is $q_{c-} = \Omega/(2 + \alpha_c)$ (Appendix~\ref{appendix:B}). We also derive an analytical expression for the classical momentum gap edge $Q_{c-}$ with the desired accuracy up to any order in $\alpha_c$ (Appendix~\ref{appendix:C}). The difference between the inverse of these critical points is
    \begin{align}
        \Omega/Q_{c-} - \Omega/q_{c-}= - \frac{1}{8} \alpha_c^2 - \frac{5}{64} \alpha_c^3 + \frac{271}{1536} \alpha_c^4 - \frac{7885}{36864} \alpha_c^5 + \frac{113531}{589824} \alpha_c^6 - \frac{1058321}{9437184} \alpha_c^7  \cdots.
    \end{align}
    Figure~\ref{fig:D1} illustrates how these analytical solutions capture this momentum shift. It can be noted that including higher-order terms in $\alpha_{c}$ yields more precise results.

\section{\,\,\,\,\,\, Inverse participation ratio and the localization-to-delocalization transition}
\label{appendix:E}
    \noindent The inverse participation ratio (IPR) serves as a reliable indicator of the localization-to-delocalization transition, complementing the role of the localization length. Formally, the IPR is defined by
\begin{equation}
\left(\mathrm{IPR} \right)_{\psi}
= 
\frac{\sum_{n=0}^{N} \bigl[\psi^*(n)\,\psi(n)\bigr]^2}
     {\bigl(\sum_{n=0}^{N} \psi^*(n)\,\psi(n)\bigr)^2},
\end{equation}
where $\psi(n)=\langle n | \psi \rangle$ is the coefficient of the wave function $\ket{\psi}$ in the orthonormal, complete basis $\{\ket{n}\}$. For a normalized state $\ket{\psi}$, the denominator simplifies to unity. The IPR reaches its maximum value of 1 if $\ket{\psi}$ occupies only one basis state $\ket{n}$. 
When $\ket{\psi}$ is fully delocalized in a space with size $N$, $\langle n |\psi\rangle \sim 1/\sqrt{N}$ and  the IPR$\sim 1/N$, approaching to zero in the thermodynamic limit ($N\rightarrow \infty$). 
    Let us specifically choose $\ket{\psi} = \ket{\psi_m}$ where $m=0,1,2,\cdots$, an eigenstate of the effective Floquet Hamiltonian, with $\ket{n}$ identified as the Floquet--photon number basis state $\ket{n,n;n}$. 
    \begin{figure}[htb!]
        \centering
        \includegraphics[width=0.75\textwidth]{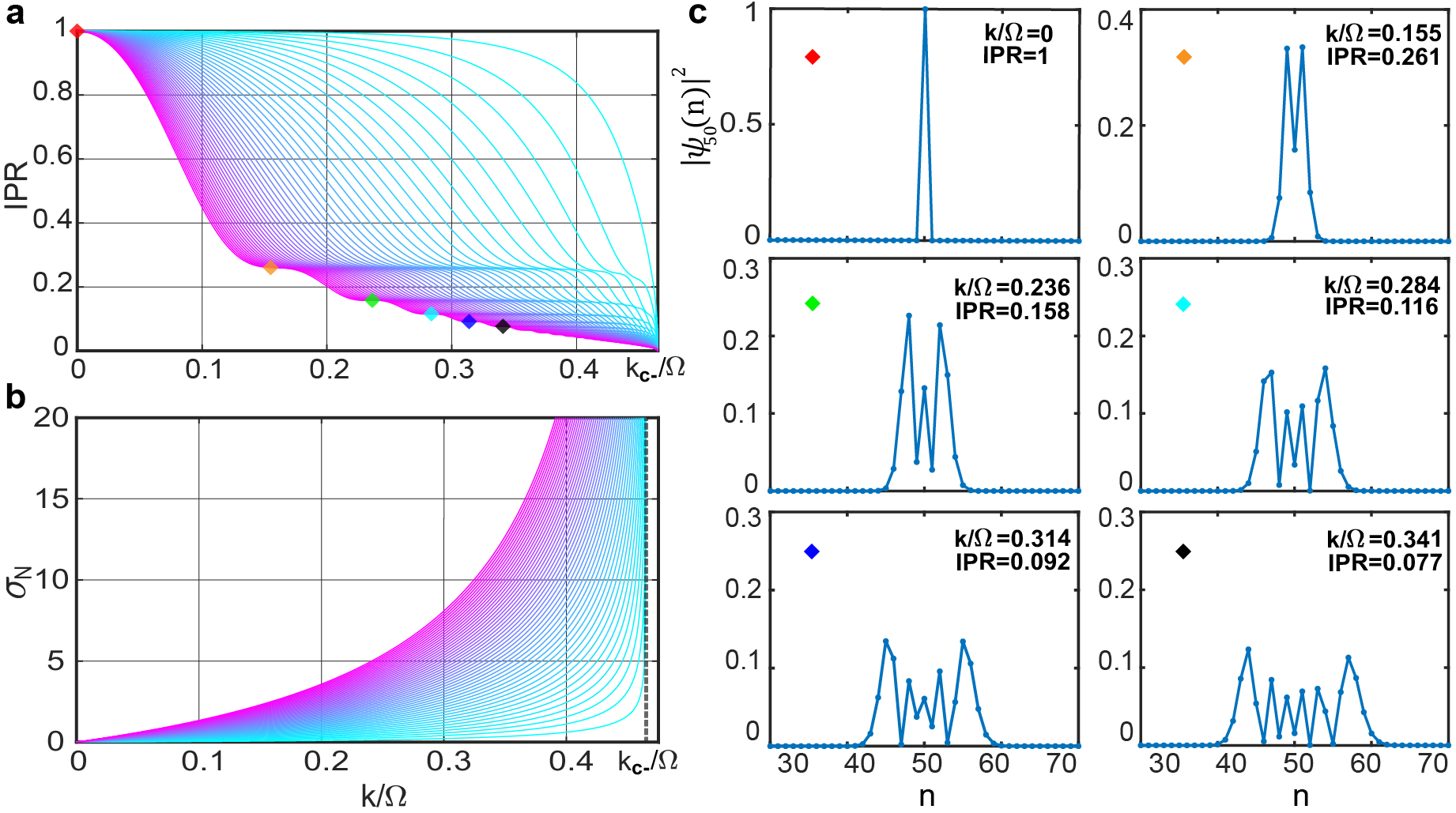}
        \caption{\label{fig:IPR} Characterizing the localization of eigenstates $\ket{\psi_m}$ of $\hat{H}^{\text{eff}}_{F}$ (with $\delta n_{k} = 0, m=0,1,2,\cdots 50$). (a) and (b) plot the IPR and the photon-number standard deviation for these eigenstates as functions of the photon momentum $k/\Omega$, for eigenvalues up to the 51th.  (c) The probability distribution $P(n)=|\langle n | \psi_m\rangle|^2$ of the 50th eigenstate $\ket{\psi_{m=50}}$ over the Floquet--photonic lattice, where $n$ abbreviates the photonic states $\{\ket{n,n;l_F=n}\}$. Computations are done at $\alpha_{c} = 0.15.$}
        \label{fig:E1}
    \end{figure}

    As $k\rightarrow k_{c-}$, the values of the IPR for each eigenstate approach zero as shown in Fig.~\ref{fig:E1}a, while the photon-number standard deviation $\left(\sigma_N\right)_{\psi_m} = \sum_{n=0}^{N} \bra{\psi_m} \hat N^2 \ket{\psi_m}-\bra{\psi_m} \hat N \ket{\psi_m}^2$ diverges, Fig.~\ref{fig:E1}b. This indicates that the localization-to-delocalization transition occurs at the critical momentum, with all eigenstates becoming delocalized. When $k \in \mathcal{K}_{\text{band}}$ (the localized region), states with larger eigenvalues tend to spread earlier. This is consistent with the scaling of hopping terms of $\hat{H}^{\text{eff}}_F$ linear in $n$.
    
    It is instructive to note that the inverse participation ratio (IPR) exhibits unusual plateaus not captured by the photon-number standard deviation $\sigma_N$. At each such plateau, the IPR ceases to decrease and instead remains nearly constant. Figure~\ref{fig:E1}c illustrates the probability distribution of the 51th eigenstate $\ket{\psi_{m=50}}$ in the Floquet-photon number basis, revealing that each new plateau emerges when the wave function acquires an additional node–antinode pair in the direction toward $n=0$, not in the direction toward $n\rightarrow \infty$ as the IPR of $\ket{\psi_{m=0}}$ does not show any plateau. Here, the antinode corresponds to a local maximum in the probability amplitude, which the IPR can discern. In contrast, $\sigma_N$ only reflects the overall spatial (photon-number) width of the wave function and thus remains insensitive to these node–antinode structures. 
    
\section{\,\,\,\,\,\, Photonic energy oscillations in the localized regime of the quantum PTC}
\label{appendix:F}

\noindent 
When a quantum state is prepared at $\ket{\psi(t=0)}=\ket{n_k=n_{-k}=l_F=0}$, one would not expect any dynamics because it corresponds to the vacuum. However, as $k\rightarrow k_{c-}$, the localization length of eigenstates diverges and the vacuum state becomes the superposition of eigenstates with different eigenenergies.  As a result, $\ket{\psi(t)}$ shows oscillating motions for $k\in\mathcal K_{\text{band}}$ in the Floquet-photonic space, and so does the photonic energy $E_{\text{ph}}(t)=\bra{\psi_t}\hat H_{\text{ph}} \ket{\psi_t}$ as shown in Fig.~\ref{fig3}b. In this section, we provide detailed analysis of the periodicity of oscillations. 
The vacuum state can be expanded in the eigenbasis of the effective Floquet Hamiltonian $\hat{H}^{\text{eff}}_F$ according to
\[
\ket{0} = \sum_m c_m \ket{\psi_m},
\]
where \( \ket{\psi_m} \) are eigenstates labeled such that the state with the smallest photon number corresponds to \( m = 0 \), and the photonic energy increases sequentially with higher \( m \).

\begin{figure}[htb!]
  \centering
    \includegraphics[width=0.9\textwidth]{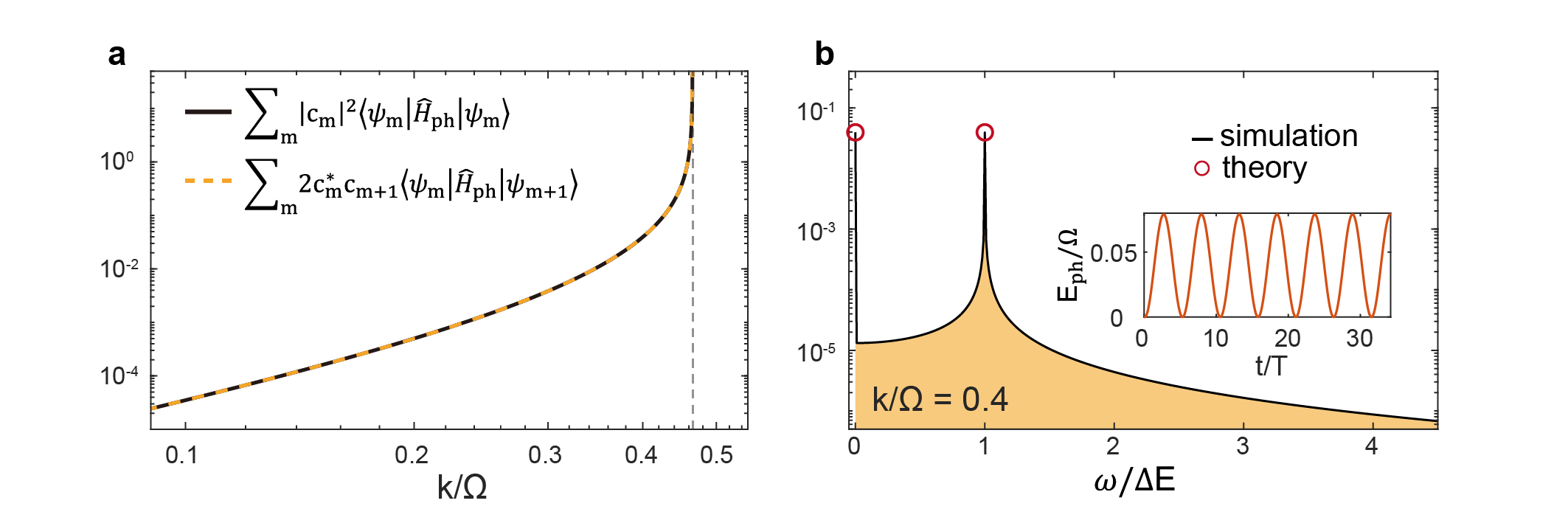}
  \caption{(a) Amplitude of the DC component $\sum_m |c_m|^2 \langle \psi_m | \hat{H}_{\text{ph}} | \psi_m \rangle$ (solid black line) and the oscillating component $ \sum_m 2c_m^* c_{m+1} \langle \psi_m | \hat{H}_{\text{ph}} | \psi_{m+1} \rangle $ (yellow dashed line) as the momentum $ k $ approaches the momentum gap edge. Here, $\delta n_k$ is set to 0. Both components exhibit divergence near the edge. (b) Frequency spectrum of $E_{\text{ph}}(t)$ for $k/\Omega = 0.4$, showing a DC peak and additional peaks at $\omega/\Delta E=1$. The black line represents simulation results, and red circles denote theoretical predictions. The inset displays the simulated time evolution of $E_{\text{ph}}(t)$, consistent with the spectrum.} \label{fig:dc_oscillating_components}
\end{figure}
The photonic energy $E_{\text{ph}}(t)$ of a system initially in the vacuum state is given by 
\begin{equation}
\begin{aligned}
   E_{\text{ph}}(t) &= \bra{\psi(t)} \hat{H}_{\text{ph}} \ket{\psi(t)} \\
    &= \left( \sum_m c_m^* e^{i E_m t} \bra{\psi_m} \right) \hat{H}_{\text{ph}} \left( \sum_{m'} c_{m'} e^{-i E_{m'} t} \ket{\psi_{m'}} \right) \\
    &= \sum_{m,{m'}} c_m^* c_{m'} e^{i (E_m - E_{m'}) t} \bra{\psi_m} \hat{H}_{\text{ph}} \ket{\psi_{m'}}.
\end{aligned} \label{eqn:Nt_in_band1}
\end{equation}
Since $\hat{H}^{\text{eff}}_{F}$ is real symmetric, its eigenstates $\ket{\psi_m}$ can be taken to have real amplitudes, making the coefficients $c_m$ real numbers. In the thermodynamic limit for $k < k_{c-}$, the energy spacing $\Delta E$ between successive eigenenergies becomes uniform, that is, $E_m = E_0 + m \,\Delta E$. Substituting this into Eq.~\eqref{eqn:Nt_in_band1}, we obtain
\begin{equation}
    E_{\text{ph}}(t) = \sum_m |c_m|^2 \bra{\psi_m} \hat{H}_{\text{ph}} \ket{\psi_m} + 2 \sum_{m} \sum_{{m'} > m} c_m^* c_{m'} \bra{\psi_m} \hat{H}_{\text{ph}} \ket{\psi_{m'}} \cos\big[(E_m - E_{m'}) t\big]. \label{eqn:Nt_in_band2}
\end{equation}
Since $E_{m'} - E_{m} = (m' - m)\,\Delta E$, we introduce $\ell = m' - m$ to rewrite the oscillatory sum:
\begin{equation}
    E_{\text{ph}}(t) = \sum_m |c_m|^2 \bra{\psi_m} \hat{H}_{\text{ph}} \ket{\psi_m} + 2 \sum_{\ell=1}^\infty \sum_m c_m^* c_{m+\ell} \bra{\psi_m} \hat{H}_{\text{ph}} \ket{\psi_{m+\ell}} \cos(\ell \Delta E t). \label{eqn:Nt_in_band3}
\end{equation}
The first term represents the time-independent average photon number (DC component), while the second term corresponds to the oscillatory components. In the localized regime, \( \bra{\psi_m} \hat{H}_{\text{ph}} \ket{\psi_{m+\ell}} \) decreases rapidly with increasing \( \ell \). As illustrated in Fig.~\ref{fig:dc_oscillating_components}a, both the DC component ($l=0$) and the oscillating term ($l=1$) grow sharply near the momentum gap edge. On the other hand, for \( \ell \geq 2 \), the terms become negligible as shown in Fig.~\ref{fig:dc_oscillating_components}b, leaving only the dominant frequency component corresponding to \( \ell = 1 \). This implies that the oscillation frequency of $E_{\text{ph}}(t)$ in the localized regime corresponds to the energy spacing \( \Delta E \) of the quantum PTC.

\section{\,\,\,\,\,\, Rabi dynamics in the quantum PTC with an initially ground-state atom}
\label{appendix:H2}
\begin{figure}[htb!]
        \centering
        \includegraphics[width=0.8\textwidth]{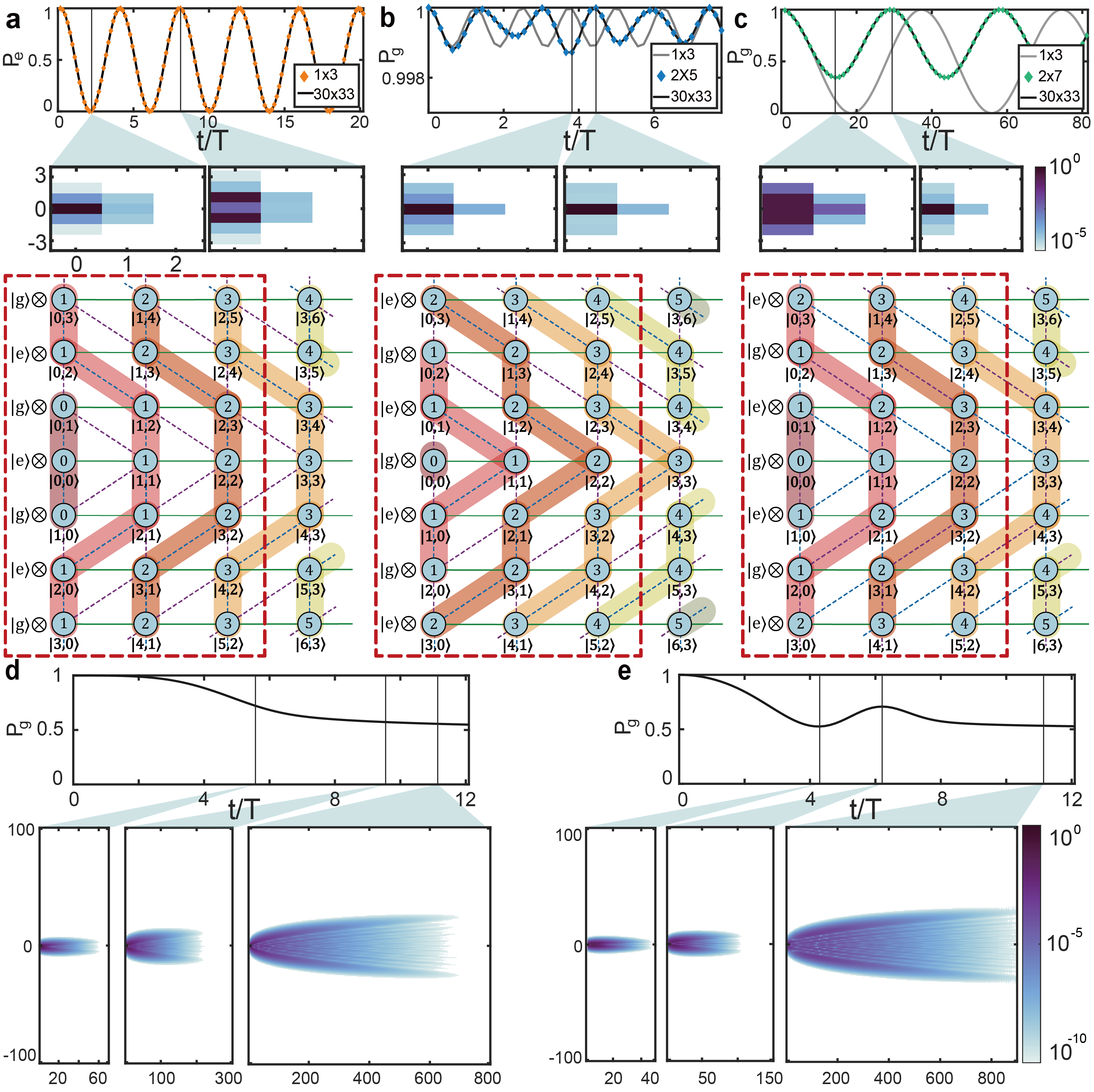}
        \caption{Dynamics of the wave function initially prepared at $\ket{n_{k},n_{-k};l_{F}} =\ket{0,0;0}$ with various atomic state conditions. (a-c) The localized regime simulations are conducted under the conditions of $\alpha_{c} = 0.15$, $k=0.215k_{c}$, and $g=0.2$. The upper panels represent the probability of excited or ground atomic state over time with different system sizes $(N_x,N_y)$. 
        The lower panels are the Hilbert space most relevant for the quantum dynamics of corresponding atom-photon hybrid systems. 
        The middle panels show the distribution of wave function in the Hilbert space (red dotted box). 
        (a) The atomic energy spacing is set to $\Delta = k$ and an initial state is prepared in $\ket{0,0;0}\otimes \ket{e}$. (b,d) The atomic energy spacing is set to $\Delta = k$ and an initial state is prepared in $\ket{0,0;0}\otimes \ket{g}$. (c,e) The atomic energy spacing is set to $\Delta = \Omega - k$ and an initial state is prepared in $\ket{0,0;0}\otimes\ket{g}$. 
        (d) A similar simulation is performed as (b) but $k=1.075k_{c}$ is used.
        (e) A similar simulation is performed as (c) but $k=1.075k_{c}$ is used.
        For the latter two cases, using $k\in\mathcal K_{\text{gap}}$, the delocalization of quantum state is observed as shown in  Fig.~\ref{fig4}.d, and the atomic state converges to $\rho_{\text{at}} {=} \frac{1}{2} (|g\rangle\langle g| + |e\rangle\langle e|)$.
        }
        \label{fig:twin}
\end{figure}
    \noindent 
    The quantum Rabi model in the quantum PTC has the three independent energy scales: (i) single photon energy $\hbar c k$, (ii) the driving frequency of the permittivity $\Omega$ generating the Floquet lattice with onsite energy $-l_F\hbar\Omega$, and (iii) the energy spacing of the lowest two atomic states $\Delta$ from $\hat H_{\text{at}}=\frac{1}{2}(\hat I + \hat \sigma_z)\Delta$. And, there are two additional coupling strengths $\alpha_c$ and $g$, which are responsible for pair creation/annihilation of photons and the interaction between atomic states and the quantum PTC, respectively. 
    
    In the main text, we set the atomic energy $\Delta =\hbar ck$ which leads the resonant condition for $k\ll k_{c-}$ and the conventional quantum Rabi dynamics is seen (Fig.~\ref{fig4}b and Fig.~\ref{fig:twin}a).  In this case, the accessible synthetic lattice sites by quantum tunneling from $\ket{0,0;0}\otimes \ket{e}$ are $\ket{0,1;0}\otimes \ket{g}$ and $\ket{1,0;0}\otimes \ket{g}$, grouped by the equipotential line around $\bold n=(0,0)$. Figure~\ref{fig:twin}a (lower panel) shows that $P_e(t)$ remains the same for choosing different system sizes: $(N_x,N_y)=(1,3)$ or $(N_x,N_y)=(30,33)$, because eigenstates are much localized on each synthetic lattice site.

    When we prepare an initial quantum state at $\ket{0,0;0}\otimes \ket g$ the nearest synthetic lattice sites in terms of onsite potential energy is drawn in Fig.~\ref{fig:twin}b (lower panel with equipotential lines). The atomic energy spacing $\Delta=k$ is still used. For $k\ll k_{c-}$, the quantum state is locked and barely show a dynamics as reflected in $P_g(t)$. For the same initial quantum state, but $k\in\mathcal K_{\text{gap}}$, as shown in Fig.~\ref{fig:twin}d it shows completely different dynamics in the synthetic space by the same reason explained in the main text: the localization-to-delocalization transition also takes place in the Floquet-photonic space combined with atomic states specified in Fig.~\ref{fig:twin}b lower panel.  The atomic state approaches to the half-half mixed state $\hat \rho_{\text{at}}=\frac{1}{2}(|g\rangle\langle g| + |e\rangle\langle e|)$ from $\hat \rho_{\text{at}}(t=0)=|g\rangle\langle g|$. The photonic energy exponentially grows over time. As a result, we reach to the claimed conclusion in the main text that irrespective of initial atomic states it converges to the half-half mixed state when $k\simeq k_{c-}$ or $k\in\mathcal K_{\text{gap}}$.

    If we tune the atomic energy spacing $\Delta = \hbar \Omega-\hbar c k$, the landscape of equipotential lines drawn in Fig.~\ref{fig:twin}c lower panel looks similar to that of Fig.~\ref{fig:twin}a lower panel, but with an initial quantum state prepared at $\ket{0,0;0}\otimes \ket g$. As a result, for $k\ll k_{c-}$ it shows the oscillation similar to the conventional Rabi model. However, the $\text{min}(P_g)$ does not reach zero (see Fig.~\ref{fig:twin}c) because of the coupling network is different from Fig.~\ref{fig:twin}a, drawn by green($\alpha_c$), blue($g$), and purple($\alpha_c g$). For $k\in\mathcal K_{\text{gap}}$, the delocalization of eigenstates leads the growth of photonic energy and the dissipation of atom to the half-half mixed state as before. Interestingly, the different landscape of equipotential line by setting $\Delta=\hbar\Omega-\hbar c k$ is reflected in the non-monotonous behavior of $P_g(t)$ shown in Fig~\ref{fig:twin}e.  

\section{\,\,\,\,\,\, Spectral analysis and atomic population dynamics in the atom–quantum PTC System}
\label{appendix:G}
\begin{figure}
  \centering
    \includegraphics[width=0.9\textwidth]{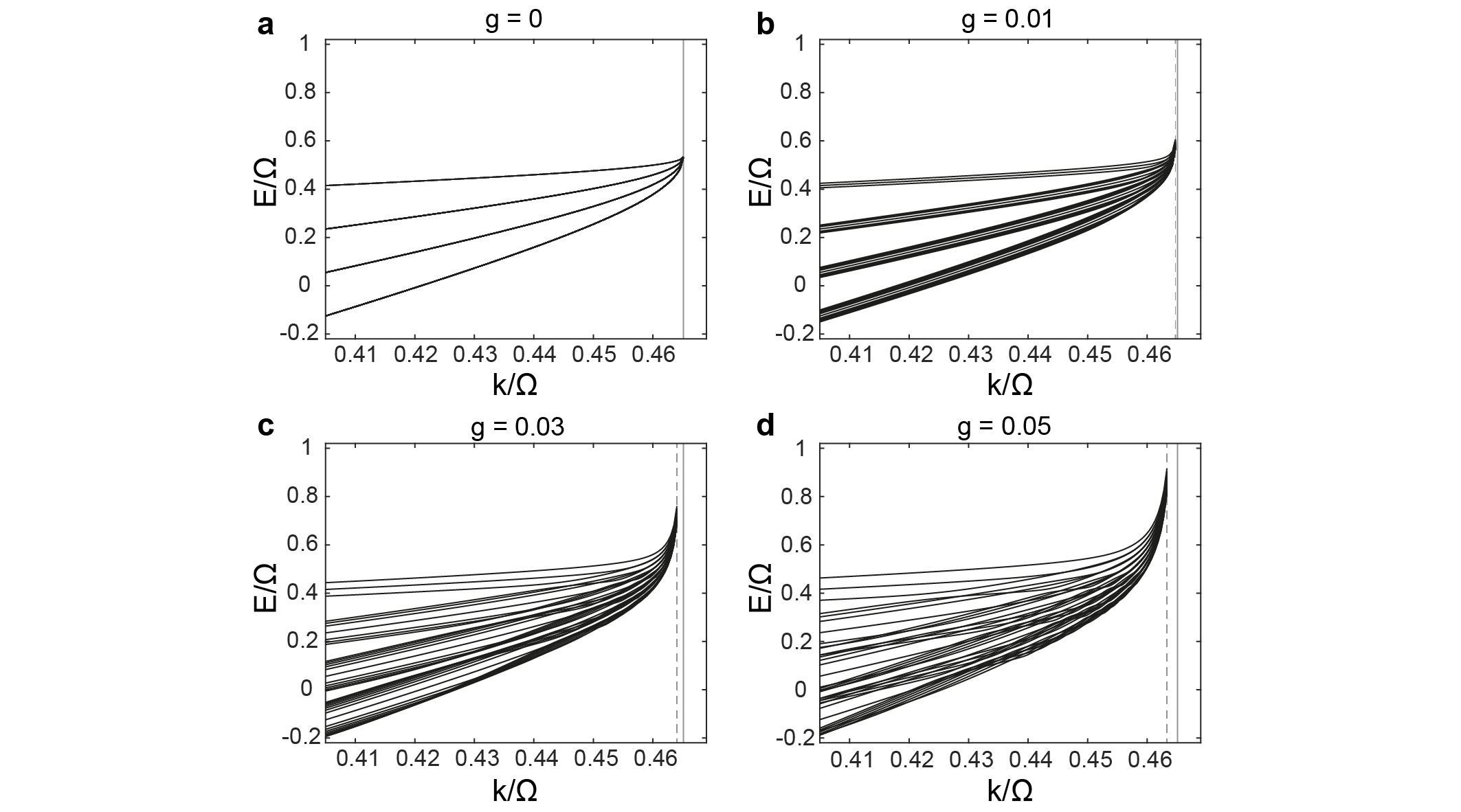}
  \caption{Quasi-eigenenergy spectra of the atom-quantum PTC coupled system, where the atomic energy is set to $E_m^{\delta n_k=1}-E_m^{\delta n_k=0}$. (a) For $g=0$, the quasi-eigenenergies of the one-dimensional wires with different momenta coupled with the atomic state are degenerate. (b-d) As the coupling strength $g$ increases, the degeneracy is lifted, causing the spectra to split. Furthermore, at larger $g$, the quasi-eigenenergy spectra converge toward smaller $k/\Omega$ and larger $E/\Omega$. Panels (b–d) illustrate this progression for $g=0.01$, $g=0.03$, and $g=0.05$.}
  \label{fig:E_diff_g}
\end{figure}

\noindent In this section, we numerically investigate how the coupling strength $g$ alters the critical momentum $k_c$. We also examine the atomic population dynamics by varying the atomic energy $E_{\text{at}}$. To evaluate the effects of $g$ on the system, we calculate the quasi-eigenenergy spectra for different values of $g$ by numerically diagonalizing the total Hamiltonian, $\hat{H} = \hat{H}^{\text{eff}}_{F} +\hat{H}_{\text{at}} + \hat{H}_{\text{int}}$. Here, the atomic energy is set to $E_m^{\delta n_k=1} - E_m^{\delta n_k=0}$, ensuring that all one-dimensional wires with different momenta $P = \delta n_k\,k$, when coupled to the atomic states (ground states for $\delta n_k = 1,3,5,\dots$ and excited states for $\delta n_k = 0,2,4,\dots$), share the same degenerate point $(k_c,E)$. When $g=0$, the quasi-eigenenergy spectrum of the atom--quantum PTC coupled system consists of degenerate bands merging at $k_c$, as illustrated in Fig.~\ref{fig:E_diff_g}(a), where the four lines correspond to $\delta n_k=0,1,2,3$, respectively. As $g$ increases, these degenerate bands begin to split. Notably, near $k_{c-}$, initially split bands remerge, causing the critical point $k_{c-}$ to shift to lower $k$, as indicated by the dotted line in Fig.~\ref{fig:E_diff_g}(b)--(d). In addition, the energy at which the spectra reconverge shifts toward higher values.

\begin{figure}
  \centering
    \includegraphics[width=0.9\textwidth]{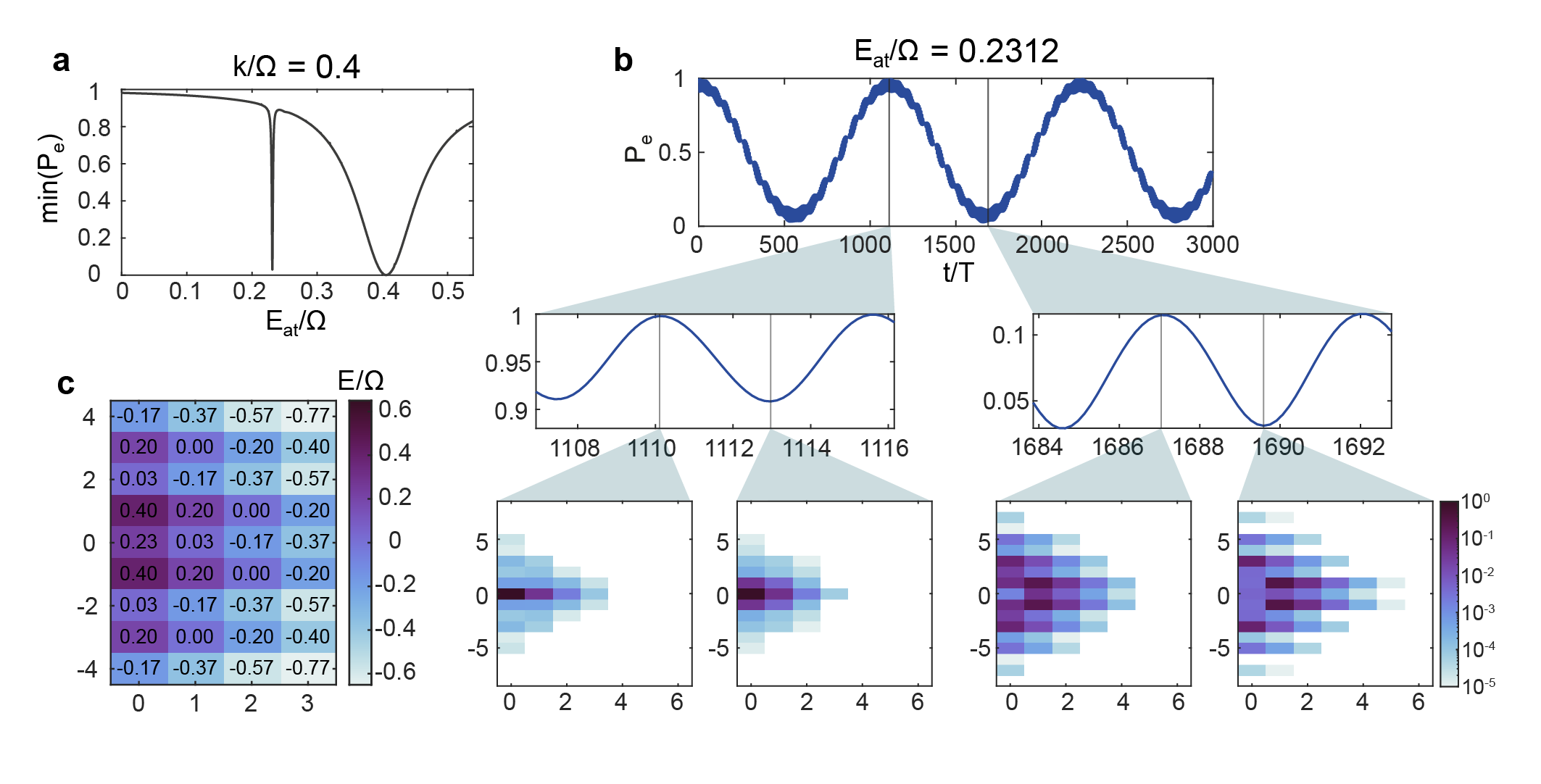}
  \caption{(a) The minimum excited-state probability min$(P_e)$, as a function of the atomic energy $E_{\text{at}}/\Omega$. An additional resonant point is observed at $E_{\text{at}}/\Omega =0.2312$ (b) The excited-state population $P_e$ as a function of time for $E_{\text{at}}/\Omega =0.2312$. The dynamics exhibit both long-period and short-period oscillations. The long-period oscillations correspond to transitions between $\ket{0,0;0}\otimes\ket{e}$ and superposition of the $\ket{1,2;-1}\otimes\ket{g}$ and $\ket{2,1;-1}\otimes\ket{g}$. The short-period oscillations involve a breathing-like motion centered around states with neighboring sites. The lower panels show $|\langle \bold n |\psi_t\rangle|^2$. (c) Onsite energies of the Floquet-photonic lattice coupled to atomic states for $E_{\text{at}}/\Omega =0.2312$ (see Fig.~\ref{fig4}a for the corresponding base states at $\bold n$). The colormap illustrates the variation of onsite energies across the lattice sites.}
  \label{fig:E_diff_Eat}
\end{figure}

\noindent We next examine the emergent resonances and population dynamics in the atom--quantum PTC coupled system, focusing on how the atomic energy $E_{\text{at}}$ affects the excited-state probability $P_e$. 
Concretely, we run the time evolution of a quantum state initially prepared at $\ket{\psi(t=0)} = \ket{n_k=0,n_{-k}=0;l_F=0}\otimes \ket{e}$ and measure the population of excited state: $P_e(t) = \text{Tr} \left[\hat \rho(t) \otimes |e\rangle \langle e| \right] $ where $\hat \rho(t) = |\psi(t)\rangle\langle \psi(t)|$ and the trace is over the Floquet-photonic basis. 
As shown in Fig.~\ref{fig:E_diff_Eat}(a), we plot the minimum value of $P_e$, $\min(P_e)$, as a function of the normalized atomic energy $E_{\text{at}}/\Omega$. 
In our simulations, $E_{\text{at}}$ is varied, and the minimum $P_e$ is extracted from the time evolution of the atomic population. A pronounced resonant dip occurs near $E_{\text{at}}/\Omega = 0.2312$. Figure~\ref{fig:E_diff_Eat}(b) displays the temporal evolution of $P_e$ at this resonance, revealing both long- and short-period oscillations. The long‐period oscillations originate from the transitions between $\ket{0,0;0}\otimes\ket{e}$ and the superposition of $\ket{1,2;-1}\otimes\ket{g}$ and $\ket{2,1;-1}\otimes\ket{g}$.  Superimposed on these slow oscillations are short‐period breathing modes, wherein the wave packet moves back and forth between adjacent lattice sites. Finally, Fig.~\ref{fig:E_diff_Eat}(c) depicts the onsite energies of the Floquet--photonic lattice coupled to the atomic states at $E_{\text{at}}/\Omega = 0.2312$.

\newpage
\end{document}